\documentclass[11pt,english]{article}
\usepackage{rotating}
\usepackage{epsfig}
\usepackage{latexsym}
\usepackage{graphicx}
\usepackage{psfrag}
\usepackage{amsmath,amssymb}
\makeatletter
\def\section{\@startsection {section}{1}{\z@}{-3.5ex plus -1ex minus
 -.2ex}{2.3ex plus .2ex}{\large\bf}}
\def\subsection{\@startsection{subsection}{2}{\z@}{-3.25ex plus -1ex
minus -.2ex}{1.5ex plus .2ex}{\normalsize\bf}}
\makeatother
\makeatletter
\def\theequation{\arabic{section}.\arabic{equation}}

\@addtoreset{equation}{section}
\renewcommand{\theequation}{\thesection.\arabic{equation}}
\makeatother

\newcommand{\captionfonts}{\small}
\makeatletter  
\long\def\@makecaption#1#2{%
  \vskip\abovecaptionskip
  \sbox\@tempboxa{{\captionfonts #1: #2}}%
  \ifdim \wd\@tempboxa >\hsize
    {\captionfonts #1: #2\par}
  \else
    \hbox to\hsize{\hfil\box\@tempboxa\hfil}%
  \fi
  \vskip\belowcaptionskip}
\makeatother   

\catcode`@=11
\def\marginnote#1{}
\newcount\hour
\newcount\minute
\newtoks\amorpm
\hour=\time\divide\hour
by60
\minute=\time{\multiply\hour by60 \global\advance\minute
by-\hour}
\edef\standardtime{{\ifnum\hour<12 \global\amorpm={am}
\else\global\amorpm={pm}\advance\hour by-12 \fi
 \ifnum\hour=0
\hour=12 \fi
 \number\hour:\ifnum\minute<10
0\fi\number\minute\the\amorpm}}
\edef\militarytime{\number\hour:\ifnum\minute<10
0\fi\number\minute}
\def\draftlabel#1{{\@bsphack\if@filesw
{\let\thepage\relax
 \xdef\@gtempa{\write\@auxout{\string
\newlabel{#1}{{\@currentlabel}{\thepage}}}}}\@gtempa
 \if@nobreak
\ifvmode\nobreak\fi\fi\fi\@esphack}
\gdef\@eqnlabel{#1}}
\def\@eqnlabel{}
\def\@vacuum{}
\def\draftmarginnote#1{\marginpar{\raggedright\scriptsize\tt#1}}
\def\draft{\oddsidemargin
0.0truein
 \def\@oddfoot{\sl preliminary draft \hfil
\rm\thepage\hfil\sl\today\quad\militarytime}
 \let\@evenfoot\@oddfoot
\overfullrule 3pt
 \let\label=\draftlabel
\let\marginnote=\draftmarginnote
\def\@eqnnum{(\theequation)\rlap{\kern\marginparsep\tt\@eqnlabel}
\global\let\@eqnlabel\@vacuum}
}
\catcode`@=12

\def\bea{\begin{eqnarray}} \def\eea{\end{eqnarray}}
\def\be{\begin{eqnarray}} \def\ee{\end{eqnarray}} \def\nn{\nonumber}
 \def\c{\hspace{-5pt}} 
 \def\ov{\overline} 
\def\ds{\displaystyle} \def\de{\partial}

\def\mc{\mathcal}

\newcommand{\promille}{%
  \relax\ifmmode\promillezeichen
        \else\leavevmode\(\mathsurround=0pt\promillezeichen\)\fi}
\newcommand{\promillezeichen}{%
  \kern-.05em%
  \raise.5ex\hbox{\the\scriptfont0 0}%
  \kern-.15em/\kern-.15em%
  \lower.25ex\hbox{\the\scriptfont0 00}}

\begin{document}

\thispagestyle{empty}

\begin{center}
\hfill UAB-FT-632 \\

\begin{center}

\vspace{1.7cm}

{\LARGE\bf A Confining Strong First-Order\\[3mm]
Electroweak Phase Transition}

\end{center}

\vspace{1.4cm}

{\bf Germano Nardini$^{\,a}$, Mariano Quir\'os$^{\,a,b}$ and Andrea Wulzer$^{\,a}$}\\

\vspace{1.2cm}

${}^a\!\!$
{\em { IFAE, Universitat Aut{\`o}noma de Barcelona,
08193 Bellaterra, Barcelona (Spain)}}

${}^b\!\!$
{\em {Instituci\`o Catalana de Recerca i Estudis Avan \c{c}ats (ICREA)}}

\end{center}

\vspace{0.8cm}

\centerline{\bf Abstract}
\vspace{2 mm}
\begin{quote}\small

In the Randall-Sundrum model where the radion is stabilized by a
Goldberger-Wise (GW) potential there is a supercooled transition from
a deconfined to a confined phase at temperatures orders of magnitude
below the typical Standard Model critical temperature. When the Higgs
is localized at the IR brane the electroweak phase transition is
delayed and becomes a strong first-order one where the Universe
expands by a few e-folds. This generates the possibility of having the
out-of-equilibrium condition required by electroweak baryogenesis in
the electroweak phase transition. We have studied numerically the
region of the GW parameter space where the theory is consistent and
the latter possibility is realized. We have found that in most of the
parameter space the nucleation temperature is so low that sphalerons
are totally inactive inside the bubbles. The condition for sphalerons
to be inactive after reheating imposes an upper bound on the reheating
temperature that is weaker for heavy Higgs bosons so that the
out-of-equilibrium condition seems to favor heavy over light
Higgses. The condition for sphalerons to be active outside the bubbles
puts an upper bound on the number of e-folds at the phase transition,
roughly consistent with the critical value required by low-scale
inflation to solve the cosmological horizon problem.

\end{quote}

\vfill

\newpage
\section{Introduction}
The Standard Model (SM) of electroweak interactions is being tested to
high accuracy at present (e.g.~Tevatron) and past (e.g.~LEP) colliders
and the physics it describes will be probed at the future LHC at
CERN. In spite of its impressive experimental successes the SM has a
number of theoretical and experimental drawbacks that make it
difficult to be considered as a fundamental theory. In particular,
from the experimental side the SM does not have a good candidate for
the Dark Matter of the Universe while on the theoretical side it
exhibits a number of problems as e.g.~the unnatural fine-tuning it
requires to accommodate the weak scale along with heavier ones, as the
Planck scale.  In order to solve some of these problems several
extensions of the Standard Model have been proposed so far. The most
popular of them, the minimal supersymmetric SM extension (MSSM) which
solves the big hierarchy problem and provides interesting candidates
to Dark Matter, shares some of the other SM problems. A completely
different solution to the hierarchy problem was proposed in
Ref.~\cite{Randall:1999ee} where the Higgs field is localized in a
four-dimensional (4D) brane inside a five-dimensional (5D) space with
a warped fifth dimension: the warping red-shifts large (Planck-sized)
scales to TeV ones, thus solving the big hierarchy problem. More
recently, warped models with matter in the bulk have been widely
discussed with the aim of finding a solution to the little hierarchy
problem as well. An incomplete list includes Higgsless
theories~\cite{Csaki:2003dt},
 in which the electroweak
symmetry is broken by boundary conditions and no Higgs scalar is
needed, and models of Gauge--Higgs Unification in warped
space~\cite{Contino:2003ve}.
The latter scenario is also interpreted, via AdS/CFT, as a calculable
version of Composite--Higgs and appears particularly promising.

An interesting feature of the SM is that it contains baryon number
violation by non-perturbative effects (sphalerons at finite
temperature) that have the capability to generate the baryon asymmetry
of the Universe at the electroweak (EW) phase transition via the
so-called EW baryogenesis (EWBG)
mechanism~\cite{Quiros:1994dr}.
Unfortunately
one of the requirements of EWBG, a strong first-order phase
transition, was soon proved not to be fulfilled in the Standard Model
 (see {\it{e.g.}} \cite{Kajantie:1995kf}),
thus preventing this appealing possibility.  While the possibility of
having a strong enough first-order phase transition is marginal in the
MSSM~\cite{Carena:1997ki} we will prove in this paper that the
Randall-Sundrum (RS) theory provides in a natural way a supercooled
first-order electroweak phase transition that can accommodate the
mechanism of EWBG provided an extra source of $CP$-violation is
generated. The main point of the mechanism is that the radion delays
the electroweak phase transition till temperatures much lower than the
electroweak one where the order parameter of the Higgs effective
potential $\phi(T)/T$ is large. We summarize in the rest of this
section the essential points of this mechanism.

At finite temperature there are two stationary solutions of the
5D gravity partition function~\cite{Creminelli:2001th}. One is the
usual RS geometry with two branes located on the UV and the IR points
respectively. The other one corresponds to AdS-Schwarzschild (AdS-S)
geometry where the IR brane is replaced by a black hole horizon. Using
the AdS-CFT correspondence both phases correspond in 4D respectively
to confined and deconfined strongly interacting gauge theory.  Because
conformal invariance is only spontaneously broken in the confined
phase the free thermal energy of the AdS-S phase is lower than that of
the RS phase and thus the latter is metastable. However after
explicitly breaking conformal invariance of the RS phase, e.g.~by the
introduction of a 5D field which creates a potential for the radion
stabilizing the distance between the UV and IR branes as in the
Goldberger-Wise (GW) model~\cite{Goldberger:1999un}, the free energy
of the confined and deconfined phases becomes equal at a given
critical temperature and for lower temperatures the phase transition
from the AdS-S to the RS phase can proceed. We have studied
numerically the process of bubble formation (of the IR brane out of
the black hole horizon) and found that~\footnote{Some preliminary
analyses of the phase transition can be found in
Refs.~\cite{Creminelli:2001th,Randall:2006py,Kaplan:2006yi}.}, though
the critical temperature lies in the sub-TeV region the actual
nucleation temperature is orders of magnitude below, essentially due
to the flatness of the radion stabilizing potential. Thus the radion
phase transition is a supercooled one and it corresponds to a strong
first-order phase transition.

On the other hand the Higgs is localized on the IR brane: it can be
either exactly localized or exponentially localized, in both cases it
corresponds to a composite object from the holographic point of
view~\footnote{The rest of the SM matter fields (fermions and gauge
bosons) can be either localized in the IR brane or they can be
propagating in the bulk of the fifth dimension.}. In the AdS-S phase,
then, the Higgs is deconfined and so from an effective theory point of
view it can be described as being in the symmetric phase $\phi=0$. In
the RS phase the Higgs field appears (confines) and so the possibility
of having $\phi\neq 0$ opens up. The radion supercooling prevents then
the Higgs phase transition to proceed at typical electroweak
temperatures. When the barrier between the deconfined and the confined
phases is sufficiently low to be overcome by a quantum jump at low
temperatures the ratio $\phi(T)/T$ can be large and the electroweak
phase transition becomes a strong first-order one. In the present
article, for simplicity, we restrict ourselves to the case of an
exactly localized Higgs. In the conclusion we will comment on how this
mechanism can be applied to other scenarios like Higgsless or
gauge-Higgs unification theories.

The contents of this article are as follows. In section~2 we describe
the two phases, RS and AdS-S, from the holographic point of view and
provide the free energy of both phases. In section~3 we add the
Goldberger-Wise bulk field in order to stabilize the radion potential
and the Higgs localized in the IR brane that provides a vacuum
expectation value (VEV) to spontaneously break the electroweak
symmetry. We identify the region in the parameter space where there is
a stable minimum fixing the radion VEV and where the back-reaction on
the RS metric is small and the theory is consistent. In section~4 the
phase transition is analyzed in some detail using both numerical
methods and analytical approximations. In most of the available
parameter space the Euclidean action is dominated by $O(4)$ symmetric
bubbles and the nucleation temperature corresponds to a few e-folds of
inflation. In section~5 we apply the results of the supercooled
electroweak phase transition to the required conditions for
EWBG. Sphalerons inside the bubbles are inactive in all cases because
the nucleation temperature is far lower than the SM phase transition
temperature.  Sphalerons have to be active outside the bubbles and
this leads to an upper bound on the number of e-folds of inflation, as
$N_e<26$, roughly consistent with the number of e-folds $(N_e\gtrsim 30)$
required to solve the cosmological horizon problem with TeV scale
inflation~\cite{Knox:1992iy}.  A second condition is obtained from
requiring sphalerons not to erase the generated baryon asymmetry after
reheating, which implies a maximal reheating
temperature. Counter-intuitively the larger the Higgs mass the weaker
this constraint becomes. This latter condition, for a given number of
e-folds of inflation, translates on upper bounds on the IR scale
$(\mu_{TeV})$ --obtained from the minimum of the GW potential-- as a
function of the Higgs mass. This leads to values of $\mu_{TeV}$ in the
TeV range at least for a moderate number of e-folds of inflation.
Again the larger the Higgs mass the weaker this condition. Finally
section~6 is devoted to our conclusion and a list of open problems.

\section{Holography of the two Phases}

The conjectured AdS/CFT correspondence~\cite{Maldacena:1997re}
relates extra--dimensional (stringy) models of gravity to $4D$ gauge theories
formulated on the boundary of the extra space. One is often interested in the
small string length limit in which the gravity side is well described by
effective (super--)gravity and, according to the correspondence, the $4D$
theory has large 't Hooft coupling. The inverse loop expansion parameter for
gravity is dual to the number $N$ of colors, so that weakly coupled
extra--dimensional models are dual to large$-N$ strongly coupled $4D$
theories. In fact we will never be able to exit the strong ('t Hooft) coupling
regime with our field--theoretical description because, unlike the more
familiar QCD case, both the confining and deconfining phases are strongly
coupled in the dual $4D$ theory we are discussing.

We will consider the Randall--Sundrum (RS) set--up~\cite{Randall:1999ee}, {\it
  i.e.} $5D$ gravity with negative cosmological constant and two
four-dimensional boundaries which are called UV and IR branes.
Holographically, this corresponds~\cite{ArkaniHamed:2000ds}
to a $4D$ conformal field theory (CFT) coupled to gravity in which the
conformal symmetry is also spontaneously broken. The spontaneous breaking is
an IR effect and its scale $\mu$ is associated to the position of the IR
brane. The presence of the UV brane, on the contrary, explicitly breaks the
conformal symmetry as it couples the theory to gravity. This occurs at an UV
scale $k=L^{-1}$, where $L$ is the AdS radius, which also represents an UV
cut--off for the CFT. At the technical level, a more precise statement of the
correspondence is as follows~\cite{Maldacena:1997re,Gubser:1999vj}. The partition
function $Z_4[M]$ of the $4D$ gauge theory formulated on the space $M$ with
metric $g$ is computed as a constrained partition function for $5D$ gravity
\be Z_4[M]\,=\,\int {\mc
D}G(x,z)_{\widehat{G}=g(x)}\ds{e^{i\,S_{grav}[G]}}\,,
\label{correspondence}
\ee
where the subscript ``$\widehat{G}=g$'' means that the $5D$ metrics $G(x,z)$
one integrates over are only those which induce a given $4D$ metric $g$ on the
UV boundary. In Eq.~(\ref{correspondence}) $S_{grav}$ is the standard $5D$
gravity action with boundaries and will be better specified below. Let us now
briefly remind what Eq.~(\ref{correspondence}) implies for the standard
interpretation of the RS model. The l.h.s. is the partition function of the
CFT calculated with a cut--off $k$, which contains not only
conformal--invariant terms but also local (non--conformal) counterterms such
as for instance an Einstein term for the $4D$ metric $g$~\cite{Gubser:1999vj}.
In the RS set--up the metrics one integrates over have no constraints at the
UV, so that the RS partition function is obtained by integrating again the
r.h.s. of Eq.~(\ref{correspondence}) on all the $4D$ metrics $g(x)$. On the
l.h.s.  this corresponds to make gravity dynamical with a $4D$ Planck scale
set by the CFT cut--off $k$. However in our context gravity is not required to
be made dynamical and all we need is Eq.~(\ref{correspondence}) in its
Euclidean version.

The thermal partition function of the $4D$ theory is indeed obtained by
considering a space $M=S^1\times T^3$ where the circle $S^1$ represents
Euclidean time and the $3$--torus $T^3$ ordinary three--space. The metric is
Euclidean, the length of the circle is $\beta=1/T$ ($T$ is the temperature)
and we will call $V$ the volume of $T^3$. At finite temperature, as noticed
in~\cite{Creminelli:2001th}, the $5D$ path integral in
Eq.~(\ref{correspondence}) has two stationary solutions at the classical
level. The first one is the Euclidean version of the standard RS geometry with
compactified time. It is
\be ds_{RS}^2\,=\,\frac{L^2}{z^2}\left(\beta^2d\tau^2\,
+\,V^{2/3}d{\vec\xi}^{\;2}+dz^2\right)\,,
\label{RS}
\ee
where the extra coordinate $z$ is in the $[z_{UV}\equiv
L,\,z_{IR}\equiv\mu^{-1}]$ interval, $\tau$ is a temporal variable with unit
period and ${\vec\xi}$ span a square $3$--torus of unitary volume. The second
solution is the so--called AdS--Schwarzschild (AdS-S) space, with metric
\be
ds_{AdS-S}^2\,=\,\frac{L^2}{z^2}\left(\beta_{h}^2
\left(1-z^4/z_{h}^4\right)d\tau^2\,+\,V^{2/3}d\vec\xi^{\;2}+
\frac{dz^2}{1-z^4/z_{h}^4}\right)\,.
\label{ADSS}
\ee
Also in this case we chose the extra coordinate $z\in[L,\,z_h]$ so
that the UV brane is located at $z_{UV}=L$, but the end of the space
now is not the IR brane, but the black hole horizon $z_h$ where the
space (\ref{ADSS}) has, in general, a conical singularity. However, by
choosing
\be	
z_h\,=\,\frac{1}{\pi} \beta_h\,,
\label{zh}
\ee
we can put the deficit angle to zero and, under this condition, the
space is completely regular and it is a true solution to the Einstein
equations. The time periodicity $\beta_h$ in Eq.~(\ref{ADSS}) is
determined by Eq.~(\ref{correspondence}) which states that at the UV
the length of the time circle must be equal to $\beta=T^{-1}$. It is
\be
\beta_{h}^2\,=\,\frac{\beta^2}2\,\left[1+\sqrt{1+4\pi^4(LT)^4}\right]\,.
\ee

The physical meaning of the two solutions is quite
simple~\cite{Creminelli:2001th}. At finite temperature, one has two
classical minima of the $5D$ action, and two disconnected
semiclassical (loop) expansions could be carried out around each of
them, leading of course to completely different physical results. From
the $4D$ dual point of view, the RS and AdS-S minima correspond to two
different phases, which we identify, respectively, with the confining
and deconfining phases of the gauge theory. It follows from
Eq.~(\ref{correspondence}) that the free--energies in the two phases
are proportional to the (effective) actions of the two gravity
solutions.  Depending on the temperature, and neglecting for the
moment the dynamics of the phase transition in an expanding Universe,
the physical minimum is the one with lowest $5D$ action or
(equivalently) $4D$ free--energy.

At the classical level one computes the free--energies by plugging the
solutions (\ref{RS}) and (\ref{ADSS}) into the action. The latter
reads
\bea
S_{grav}^E&=&-4M^3 \left[\ \int_{\mc
M}\left(R+12\,k^2\right)\,+\,\int_{\de{\mc M}_{UV}}2K_{UV}\right.\,\nn \\
&+&\left.\int_{\de{\mc M}_{IR}}\left(2K_{IR}+6k\right)\right]\,,
\label{action}
\eea
where $R$ is the $5D$ curvature scalar, $M\sim M_P$ the $5D$ Planck
mass and $K_{UV(IR)}$ the extrinsic curvatures of the two boundaries
$\de{\mc M}_{UV(IR)}$. Following the conventions
of~\cite{Kraus:1999it}, in Gaussian normal coordinates
\be
K\,=\,\gamma^{ab}\,K_{ab}\,=\,\frac12 \gamma^{ab}\de_\eta\gamma_{ab}\,,
\ee
where $\eta$ is the coordinate normal to the boundary and $\gamma$ the induced
metric. The equations of motion associated to the action (\ref{action}) are
the $5D$ Einstein equations in the bulk plus the Israel junction condition
\be
K_{ab}^{IR}\,=\,-k\gamma_{ab}^{IR}\,,
\label{israel}
\ee
at the IR brane only. One has no junction condition at the UV because the UV
metric is fixed in (\ref{correspondence}) and should not be varied when
working out the equations of motions from the action.

While both spaces (\ref{RS}) and (\ref{ADSS}) solve the bulk Einstein
equations only the RS geometry solves the Israel junction condition
(\ref{israel}), thanks to the fine--tuned choice we did in
Eq.~(\ref{action}) for the IR brane tension. On the contrary the AdS-S
metric (\ref{ADSS}) does not solve Eq.~(\ref{israel}), meaning that an
IR brane cannot be put in that geometry. Thus strictly speaking both
spaces (\ref{RS}) and (\ref{ADSS}) are not solutions of the same
gravity action.  In fact for the AdS-S to be a solution the IR
boundary terms (last terms of Eq.~(\ref{action})) should be dropped.
One might think however that the IR brane is hidden by the black hole
horizon in the AdS-S space so that the IR terms of the action do not
effectively contribute. More precisely one should think of the IR
brane in RS as a remnant of a more complicated solitonic--like
geometry which solves the equations of motions of a more fundamental
underlying theory of gravity.  The same theory should also allow for a
second solution, without the brane, which would correspond to the
AdS-S space.

The free--energy of the two phases is readily computed. By plugging
Eq.~(\ref{RS}) into Eq.~(\ref{action}) we get for the confining (RS) phase
\be
F_{RS}\,=\,\frac1{\beta V}\,S_{grav}^E\left[RS\right]\,=\,-24(ML)^3k^4\,,
\label{frs}
\ee
which is just the usual RS bulk cosmological constant which we could
have canceled by adding a suitable UV brane tension term to
Eq.~(\ref{action}). For the deconfined phase we must remove the IR
terms from Eq.~(\ref{action}) and plug Eq.~(\ref{ADSS}). One gets
\bea
F_{AdS-S}\,&=&\,\frac1{\beta V}\,S_{grav}^E\left[AdS-S\right]\,
=\,-8(ML)^3k^4\left[\frac{\beta}{\beta_h}+2\frac{\beta_h}{\beta}\right]\,\nn\\
&=&\,\left[-24 k^4-4\pi^4 T^4\left(1+
\mathcal O\left(\frac{\pi^4T^4}{k^4}\right)\right)\,\right](ML)^3 ~,
\label{fadss}
\eea
where an expansion for $\pi T\ll k$ has been performed.

As a general
rule of AdS/CFT, the loop expansion on the gravity side corresponds to
a large number of colors (large--$N$) expansion in the gauge
theory. The squared coupling constant for $5D$ gravity is $1/(ML)^3$
and naive dimensional analysis  suggests that the expansion
parameter is given by the squared coupling divided by a loop
factor. The precise AdS/CFT relation~\footnote{The relation
(\ref{Nex}) can be changed by numerical factors depending on the
precise theory. We will use Eq.~(\ref{Nex}) as a definition.}
\be
\frac1{N^2}\,=\,\frac{(ML)^{-3}}{16\pi^2}\,,
\label{Nex}
\ee
perfectly matches this interpretation. Let us now compare the above
equation with the free--energies (\ref{frs}) and (\ref{fadss}). The RS
phase is confining and indeed $F_{RS}$ only contains a vacuum energy
term to which all the $N^2(-1)$ confined gluons can contribute. The
vacuum energy scale is fixed by the UV cut--off $k$ of the $4D$
theory. The AdS-S phase, on the contrary, is deconfining and the
$N^2(-1)$ massless gluons are present in the plasma everyone giving
its standard $\sim T^4$ contribution. The vacuum energy, which is the
same in both phases, can be always shifted to zero by adding an UV
brane tension term to the action (\ref{action}).

\section{Adding the Golberger--Wise and Higgs fields}

Equations (\ref{frs}) and (\ref{fadss}) immediately uncover a problem: at any
non--zero temperature $F_{AdS-S}<F_{RS}$ so that the RS space is metastable
and no phase transition can ever occur from the AdS-S to the RS phase. This is
because the conformal symmetry is only spontaneously broken in the RS phase
and it is completely unbroken in the AdS-S one. Being spontaneous, the
conformal breaking indeed leads to a massless Goldstone boson, the radion,
which arises in $5D$ as a modulus of the RS geometry corresponding to the
distance among the two branes~\cite{ArkaniHamed:2000ds}.
Having fixed the location of the UV brane, the radion field controls the
position $\mu^{-1}$ of the IR brane. Given that the radion is a Goldstone, it
cannot have a potential and this is why it does not contribute to the vacuum
energy and then to the free--energy (\ref{frs}) at leading order in the
large--$N$ expansion.  Since the radion potential is flat and $\mu$
undetermined, the temperature is the only dimensionful quantity and there is
no other scale the temperature can be compared with (the cut--off $k$ is not
relevant here). Stated in another way there is no scale which could fix the
temperature of transition $T_c$. In fact starting in the AdS-S phase and
lowering the temperature the system would remain in that phase forever. The
same occurs for the first holographic deconfining phase transition discussed
by Witten~\cite{Witten:1998zw} in the case of unbroken conformal symmetry.

\subsection{The Golberger--Wise field}
Independently of the above considerations an explicit breaking of conformal
symmetry is required anyhow to avoid the presence of an exactly massless
scalar. The so--called Goldberger--Wise (GW)
mechanism~\cite{Goldberger:1999uk} permits to stabilize the distance among the
branes and give a mass (and a potential) to the radion by simply adding a $5D$
scalar field $\Phi$. The Euclidean action is
\be
S_{GW}^E\,=\,\int_{\mc
M}\left[G^{MN}\de_M\Phi\de_N\Phi\,+\,m^2\,\Phi^2\right]+\int_{\de{\mc
M}_{IR}}\left({\mc L}_{IR}\left(\Phi\right)\,+\,k^4\delta
T_1\right)\,,
\label{gwaction}
\ee
where $\delta T_1$ amounts to a correction to the IR brane tension in
Eq.~(\ref{action}). We do not need to specify the IR Lagrangian ${\mc L}_{IR}$
since it only plays the role (as in~\cite{Goldberger:1999uk}) of enforcing the
IR boundary condition
\be
\Phi(x,z_{IR})\,=\,k^{3/2}\,v_1\,.
\label{gwir}
\ee

Holographically the GW field
corresponds~\cite{ArkaniHamed:2000ds}
to an explicit UV
breaking of the conformal symmetry due to an operator ${\mc O}$ of conformal
dimension $d=4+\epsilon$ where
\be
\epsilon\,=\,\sqrt{4\,+\,m^2L^2}\,-\,2\,\simeq\,\frac{m^2L^2}4\,,
\ee
and we have assumed $m^2\ll k^2$ to perform the expansion. In AdS space the
squared mass of a scalar can be positive or negative and then $\epsilon$ can
have both signs. To have a solution of the hierarchy problem $|\epsilon|$ must
be small ($\lesssim 1/10$) and we are going to assume this in the following.
Adding the GW term (\ref{gwaction}) to the action modifies the correspondence
of Eq.~(\ref{correspondence}) in a very simple way. On the l.h.s. one has the
partition function of a different $4D$ theory which gets modified by the
introduction of the term
\be
\frac{\lambda}{k^{\epsilon+3/2}}\,\widehat\Phi(x)\,{\mc O}\,,
\label{GWoperator}
\ee
to the $4D$ Lagrangian. In the above equation $\lambda$ is a dimensionless
coupling, $k$ is the UV cut--off and $\widehat\Phi(x)$ a non-dynamical $4D$
source field. On the r.h.s. of Eq.~(\ref{correspondence}) we now have the new
$5D$ field $\Phi$ to integrate over. As for the metric one has to consider a
constrained path integral. The value of $\Phi$ on the UV brane is
\be
\Phi(x,z_{UV})\,=\widehat\Phi\,\equiv\,k^{3/2}\,v_0\,,
\label{gwuv}
\ee
where we fixed the source $\widehat\Phi$ to a constant value and then
converted the operator (\ref{GWoperator}) into a deformation of the CFT. One
could also have made $\widehat\Phi$ dynamical, and added a Lagrangian for
$\Phi$ at the UV as in~\cite{Goldberger:1999uk}. The role of those terms, as
for ${\mc L}_{IR}$, would simply have been to enforce the boundary condition
(\ref{gwuv}) and hence, in practice, to convert $\widehat\Phi(x)$ back into a
non--dynamical field. The two approaches are then completely equivalent.

To find the background configuration in the presence of the GW scalar one has
to solve the coupled Einstein and Klein--Gordon equations which come from the
total action $S_{grav}^E+S_{GW}^E$ and impose the boundary conditions
(\ref{gwir}) and (\ref{gwuv}). This is in general a non--trivial task and
exact analytic solutions can only be obtained in some particular
cases~\cite{DeWolfe:1999cp}. The problem greatly simplifies if we can neglect
the backreaction of the GW scalar on the $5D$ metric. In this case one simply
has to solve the Klein--Gordon equation on the corresponding unperturbed
gravitational background, provided by Eqs.~(\ref{RS}) and (\ref{ADSS}), and
this provides an approximate solution of the original coupled equations.  In
the RS case the actual value of $\mu$ (i.e.~the position $\mu^{-1}$ of the IR
brane) is the one which minimizes the value of the action on the corresponding
solution. By plugging the solution into the action one gets, in addition to
the vacuum energy term of Eq.~(\ref{frs}), a potential $V_{GW}(\mu)$ for the
radion whose minimum fixes the value of the (stabilized) modulus. The result
is~\cite{Goldberger:1999uk}
\be
V_{GW}(\mu)\,=\,\epsilon v_{0}^2k^4\,+\,\mu^4\left[(4+2\epsilon)
(v_1-v_0(\mu/k)^\epsilon)^2\,-\,\epsilon v_{1}^2\,+\,\delta T_1\right]\,,
\label{gwpot}
\ee
in the very good approximation of neglecting $(\mu/k)^4\ll1$. 

For the potential (\ref{gwpot}) to have a (non--trivial) global
minimum one must require
\bea
&&\textrm{for}\;\epsilon\,>\,0\,,\;\;\;\;\;\; \frac{\delta T_1}{v_{1}^2}\,<\,
\epsilon\,,\nn\\
&&\textrm{for}\;\epsilon\,<\,0\,,\;\;\;\;\;\; -(4+\epsilon)\,<\,
\frac{\delta T_1}{v_{1}^2}\,<\,\epsilon\,.
\label{conscond}
\eea
If the previous conditions are satisfied the global minimum of $V_{GW}(\mu)$
is located at
\be
\mu_{TeV}\,=\,k\left(\frac{v_1}{v_0}\right)^{1/\epsilon}X^{1/\epsilon}
\,,\;\;\;\;
\textrm{where}\;\;\;\;X\,=\,\frac{4+\epsilon+\textrm{sign}(\epsilon)
\sqrt{\epsilon(4+\epsilon)-4\frac{\delta T_1}{v_{1}^2}}}{4+2\epsilon}\,.
\label{mutev}
\ee
It is also useful to have a formula for the value of the potential at
its minimum. It is
\be V(\mu_{TeV})\,=\,\epsilon
v_{0}^2k^4\,+\,\mu_{TeV}^4\,v_{1}^2\,\frac{\epsilon}{2+\epsilon}
\left[\frac{\delta
T_1}{v_{1}^2}-\textrm{sign}(\epsilon)\sqrt{\epsilon(4+\epsilon)-4\frac{\delta
T_1}{v_{1}^2}}\right]\,.  \ee

Equation (\ref{mutev}) shows how the GW mechanism provides a solution to the
Hierarchy Problem. Since $k$ is associated to the Planck scale and $\mu_{TeV}$
to the Electroweak ($TeV$) one, an enormous fine--tuning is expected ``a
priori'' in order to get $\mu_{TeV}$ out of microscopic parameters of order
$k$. However by making a sensible choice of $|\epsilon|$ and $v_1/v_0$ not too
far from $1$ Eq.~(\ref{mutev}) can easily give rise to the required $16$
orders of magnitude among $\mu_{TeV}$ and $k$.
Solving the Hierarchy Problem is the greatest success of this class of models,
and considering too high values of $|\epsilon|$ would spoil this success. If
we at most allow for a two orders of magnitude hierarchy among $v_1$ and $v_0$
we get the upper bound:
\be
|\epsilon|\,\lesssim\,\frac18\,.
\ee 
For $|\epsilon|=1/20$, which is the reference value we will mostly use in the
following, $v_1/v_0\sim10^{\textrm{sign}(\epsilon) 4/5}$.

For the above manipulations to make sense the GW field must provide a small
backreaction on the metric. To ensure this condition, the contribution of the
GW field to the energy--momentum tensor must be compared with the one coming
from the bulk cosmological constant. This leads to the constraints
\bea
&&\frac{\pi^2v_{1}^2}{N^2}\,<\,\frac{3}{|\epsilon(4+\epsilon)|+
\epsilon^2}\frac1{X^2}\left(\frac{\mu}{k}\right)^{2\epsilon}\,,\nn\\
&&\frac{\pi^2v_{1}^2}{N^2}\,<\,\frac{3}{|\epsilon(4+\epsilon)|+
\epsilon(4+\epsilon)-4\frac{\delta
T_1}{v_{1}^2}}\,,
\label{bac1}
\eea
where the $|\epsilon(4+\epsilon)|$ term in the denominators comes from the
fact that, depending on the sign of $\epsilon$, the stronger constraint arises
from the $T_{\mu\nu}$ or $T_{zz}$ components of the $5D$ energy--momentum
tensor $T_{MN}$. The small back--reaction condition also implies another
constraint, which was considered in~\cite{Goldberger:1999uk} and ignored
in~\cite{Randall:2006py}. It comes from comparing the RS brane tension in
Eq.~(\ref{action}) with the $\delta T_1$ term of Eq.~(\ref{gwaction}).
Remembering that the latter has been treated as a correction, we must impose
\be
|\delta T_1|\,<\,24(ML)^3\,=\,\frac{3}{2\pi^2}N^2\,.
\label{bacdt}
\ee

Depending on the sign of $\epsilon$ significantly different constraints on
$v_1/N$ arise. When $\epsilon$ is positive $(\mu/k)^{2\epsilon}$ is small and
the first bound in Eq.~(\ref{bac1}) is the relevant one. The two bounds could
only be comparable for very small values of $\epsilon$. In the following we
will however mainly be interested in negative $\epsilon$. In this case the
first bound in Eq.~(\ref{bac1}) is weak and the second bound becomes the
relevant one.  As for the latter $v_{1}^2$ drops from the second line of
Eq.~(\ref{bac1}) one gets, noticing that $\delta T_1<0$
\be |\delta
T_1|\,<\,12(ML)^3\,=\,\frac{3}{4\pi^2}N^2\,,
\label{bacdt1}
\ee
which is a factor of $2$ stronger than Eq.~(\ref{bacdt}). Considering
Eq.~(\ref{conscond}), Eq.~(\ref{bacdt1}) can be converted into a bound on
$v_1$ as
\be
\frac{\pi^2v_{1}^2}{N^2}\,<\,\frac{3}{4|\epsilon|}\,.
\label{bacfinal}
\ee
The above condition is always stronger than the one coming from the first
line of Eq.~(\ref{bac1}). Indeed, $(\mu/k)^{2\epsilon}>1$ and it is possible
to show, using Eq.~(\ref{conscond}), that $X$ is smaller than one. We can then
conclude that for $\epsilon<0$ Eq.~(\ref{bacdt1}) is the only
small--backreaction condition, although one should however account for the
consistency conditions in Eq.~(\ref{conscond}).

Focusing on the negative $\epsilon$ case it is convenient to define
the variables
\be 
\theta=\frac43\frac{\pi^2|\delta
T_1|}{N^2}\,,\;\;\;\;\;\nu=\frac43 \frac{\pi^2
|\epsilon|v_1^2}{N^2}\,. 
\label{deftv}
\ee 
The perturbativity constraint (\ref{bacdt1}) and the consistency conditions
(\ref{conscond}) can respectively be expressed as
\be
\theta<1\,,\;\;\;\;\;\nu<\theta<\frac{4-|\epsilon|}{|\epsilon|}\nu\,.
\label{perturbativity}
\ee 
As far as $\epsilon$ is small, the last inequality is substantially irrelevant
and all our constraints can be expressed in terms of two parameters only.

In order to discuss how the GW field affects the deconfined (AdS-S) phase,
following ~\cite{Creminelli:2001th}, we have to construct an ``off--shell''
version of the AdS-S geometry (\ref{ADSS}) by defining a ``Hawking
temperature'' $T_h\equiv1/(\pi z_h)$ and relaxing the regularity condition
(\ref{zh}). A $4D$ field $T_h(x)$ is associated to the parameter $T_h$. Very
much like the radion field $\mu(x)$ controls the position $\mu^{-1}$ of the
$IR$ brane in the RS phase, $T_h(x)$ controls the position $z_h$ of the AdS-S
black hole horizon. Differently from the radion, however, $T_h$ has a
potential even in the absence of the GW field. Having detuned the condition
(\ref{zh}), indeed, the space develops a conical singularity which, once
regularized with a suitable spherical cap, gives rise to a potential for $T_h$
with a minimum at $T_h={\beta_h}^{-1} (\simeq T$ for $T\ll k$). Taking also
into account the GW contribution, the $T_h$ potential [{\it{i.e.}} the
deconfined (AdS-S) phase free--energy] is
\be 
F_{d}(T_h,T)=E_0\,+\,\frac{3\pi^2 N^2}4 T_h^4\,-\,\pi^2 N^2 T
T_h^3\,+\,\frac{3\pi^2 N^2}8 \nu X^2\left(\frac{\mu_{TeV}}{\pi
T_h}\right)^{2|\epsilon|}T_h^4\,,
\label{fadss1}
\ee
where we subtracted the vacuum energy term
\be
E_0=-V_{GW}(\mu_{TeV})\,,
\ee
in order to cancel the cosmological constant term [Eq.~(\ref{gwpot}) at the
minimum] of the confined (RS) phase. At small $|\epsilon|$ the above formulas
reduce to
\be
F_{d}(T_h,T)\simeq E_0\,+\,\pi^2
N^2\left[\frac34\left(1+\frac\nu{2}\right) T_h^4\,-\, T
T_h^3\right]\,, 
\ee 
with 
\be
E_0=\frac{3N^2}{4\pi^2}\sqrt{(\theta\nu-\nu^2)|\epsilon|}\mu_{TeV}^4\,.
\label{vacuumE}
\ee 
Notice that the above expansion holds for $|\epsilon|^{1/2}\ll 1$ and fails
for $\theta\gg \nu$. In this approximation $F_d$ has a minimum at
\be 
T_h^{min}=\frac{T}{1+\nu/2}\,, 
\ee 
and the free--energy at the minimum is
\be F_{d}^{min}(T)=E_0\,-\,\frac{\pi^2
N^2}4\frac{T^4}{(1+\nu/2)^3}\,.
\label{fadssm}
\ee

\begin{figure}[htb]
\psfrag{t}[][bl]{$\theta$}\psfrag{n}[][l]{$\nu$}
%
\vspace{.5cm}
\begin{center}
\epsfig{file=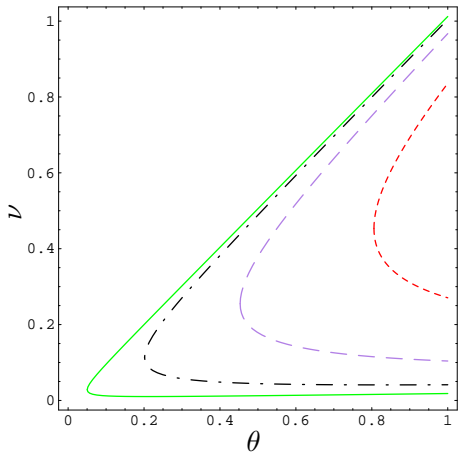,width=0.4\textwidth}
\end{center}
\caption{Level curves of $m_r/m_{KK}$ equals to 0.25 (red
 short-dashed line), 0.19 (purple
 dashed), 0.13 (black dot-dashed) and 0.06 ( 
 green solid)   in the plane
 ($\theta$,$\nu$). We have fixed $\epsilon=-1/20$. }
\label{rm}
\end{figure}

The introduction of the $T_h(x)$ field associated to the horizon position also
permits to develop an intuitive understanding of how tunneling among the two
vacua occurs~\cite{Creminelli:2001th}: starting in the AdS-S phase at
$T_h=T_h^{min}$ the horizon starts to move away from the $UV$ brane since
$T_h$ reaches $0$. At this point the horizon has disappeared and the metric
(\ref{ADSS}) reduces to the one of the infinite $AdS_5$ space, which coincides
with the RS space (\ref{RS}) for $\mu=0$. The $IR$ brane now appears from
infinity and $\mu$ grows from zero to its equilibrium value $\mu_{TeV}$. The
radion, in this schematization, is the only field of the RS phase which
undergoes a variation during the tunneling process and it is only possible to
justify this approximation (at a not too high temperature compared with
$\mu_{TeV}$) if the RS KK gravitons are significantly heavier than the radion.
In this case it is energetically expensive for the KK fields to move and the
transition can be studied in an effective theory for the radion alone. With
the aim of checking that indeed $m_r<m_{KK}\approx\pi\mu_{TeV}$ in all our
allowed parameter space, and for completeness, we report here the radion
Lagrangian~\cite{Csaki:1999mp,ArkaniHamed:2000ds,Goldberger:1999un} 
\be {\mc
  L}_r\,=\,\frac{3N^2}{2\pi^2}\frac{1}{2}(\partial_\nu\mu)^2\,-\,V_{GW}(\mu)\,
  =\,\frac{3N^2}{2\pi^2}\frac{1}{2}\left[(\partial_\nu\mu)^2\,
-\,m_{r}^2\mu^2+\cdots\right]\,,
  \ee
where 
\bea m_r^2 \,&=&\, 2(2+\epsilon)\nu \left(4 X +4\epsilon X
  -2\epsilon X^2- 4 X^2 \right) \mu_{TeV}^2
\nonumber \\
&\simeq& 8 \sqrt{-\epsilon(\theta-\nu)\nu}\,\mu_{TeV}^2\,.  
\eea
Level curves of $m_r/m_{KK}$ in the region of
parameters allowed by Eq.~(\ref{perturbativity}) are shown in Fig.~\ref{rm}, 
and we see that the radion is
always sufficiently light for our approach to the transition to be meaningful.

Introducing the SM fields, as we will do in the following, does not
dramatically change the picture of the transition, we will however
have to include a second light scalar, the Higgs field, on the RS
side. The transition becomes, in principle, a problem of tunneling in
a two--dimensional field space, which is numerically much more
complicated than the standard one--dimensional one. We will argue,
however, the Higgs contribution to the bounce to be negligible and
studying the one--dimensional radion tunneling will be sufficient for
our purposes.

We conclude this section by observing that small--backreaction is not
at all necessary for the GW mechanism to work, as shown
in~\cite{DeWolfe:1999cp}. As long as $|\epsilon|$ is sufficiently
small the GW mechanism will still provide a solution to the Hierarchy
Problem even in the presence of a significant backreaction. From this
point of view the small backreaction conditions we impose only
have a {\emph{technical}} motivation, as they ensure the consistency
of our manipulations. All calculations done in this section could in
principle be repeated taking backreaction into account. One should
consider the coupled Einstein and Klein--Gordon equations, find the
two solutions and substitute them back into the action. This would
lead to exact expressions for the free energies of the two phases at
leading order in $N$. For what concerns the dynamics of the phase
transition, on the contrary small backreaction is more than a
technical assumption. The point is that, as shown in Fig.~\ref{rm}
this condition ensures the radion to be light.  If this were not the
case, as previously discussed, the schematization of the transition by
the radion tunneling would be meaningless and one would need to go to
the true extra--dimensional gravitational instanton which connects the
two vacua. This is most likely beyond the reach of our
field--theoretical description of the two phases, as one should face
the problem of describing in some way the process of the IR brane
hitting the horizon and disappearing into the black hole.

\subsection{The Higgs field}
Let us now introduce the Higgs field in our theory. As in the original
RS scenario we take the Higgs doublet $\bar h$ to be localized at the
IR brane.  The (Euclidean) action is then~\footnote{We are assuming
here for simplicity that there is no Higgs-radion mixing from a term
like $\xi \int d^4x\sqrt{\gamma^{IR}} R(\gamma^{IR})\bar h^\dagger
\bar h$, i.e.~that $\xi\simeq 0$. }
\bea 
S_{SM}^E\,&=&\,\int_{\de{\mc M}_{IR}}{\mc L}_{SM}^E\,=\,\int
 d^4x\,\sqrt{\gamma^{IR}}\,{\mc L}_{SM}^E ~,
\label{ssm}
\eea
where the Higgs doublet $\bar h$ has a tree-level potential
\be
V_0(\bar h,\bar v)
=\,\lambda\left(\bar{h}^\dag \bar{h}-\frac{1}{2}\bar{v}^2\right)^2\,.
\label{vsm0}
\ee
For the rest of SM fields we will not consider a particular
scenario although we will assume that gauge bosons and (at least part
of) the SM fermions~\footnote{Some fermions could also be localized in
the IR brane as the Higgs fields.} are propagating in the bulk. We
will also assume that for low enough energies (where all
Kaluza-Klein excitations are decoupled) the zero modes are effectively 
described by the SM.

Using now the RS metric (\ref{RS}) one can write the action as
\be
S_{SM}^E =\int d^4x\left\{ 
\left(\frac{\mu}{\mu_{TeV}}\right)^2\left|D_\mu h\right|^2+
\left(\frac{\mu}{\mu_{TeV}}\right)^4 V_0(h,v)+\cdots \right\}
\label{ssm1}
\ee
where the Higgs field is redefined as
\be
h=\frac{\mu_{TeV}}{k}\, \bar h,\ 
\left[v=\frac{\mu_{TeV}}{k}\, \bar v\simeq 246\ GeV\right].
\label{red}
\ee
Notice that Higgs fields and mass parameters (or vacuum
expectation values), are red-shifted with a power of the factor
$\mu_{TeV}/k$ according to their dimension.

The IR brane terms only contribute to the RS phase and hence, in the
present model, the Higgs field is not present at all in AdS-S. The
holographic reason for this is that the Higgs particles~\footnote{As
well as other particles possibly localized on the IR brane.} are
entirely composite objects which cannot be present in the deconfined
high--temperature plasma.

Light SM fields, on the contrary, are most likely fundamental particles. 
Including their contribution the AdS-S free energy
becomes

\begin{equation}
F_d=E_0\,-\,\frac{\pi^2
N^2}4\frac{T^4}{(1+\nu/2)^3}-\frac{\pi^2}{90} g^*_d\, T^4   ~,
\label{ultima}
\end{equation}
while in the RS phase we have
\be
F_c(\mu,\phi_c,T)=V_{RS}(\mu,\phi_c,T)\,
+\,E_0\,-\,\frac{\pi^2}{90}g_c^*\,T^4\,,
\label{frs1}
\ee
where $g^*_c\sim 100$ ($g_d^*$) is the effective number of light
SM degrees of freedom in the RS (AdS-S) phase, $\phi_c\equiv \sqrt{2} h^0$
and $V_{RS}(\mu,\phi_c,T)$ indicates the radion-Higgs potential. At
the tree-level it reads
\begin{equation}
V_{RS}(\mu,\phi_c)=V_{GW}(\mu)+\left(\frac{\mu}{\mu_{TeV}}\right)^4
\frac{\lambda}{4} (\phi_c^2-v^2)^2 ~.
\label{totalpot}
\end{equation}
As we can see from Eq.~(\ref{totalpot}) the typical size of the SM
potential is $V_{SM}\sim v^4$ while that of the GW potential is much
larger, i.e.~$V_{GW}\sim \mu_{TeV}^4$.  Therefore the total potential
\begin{equation}
V_{RS}=V_{GW}\left[1+\mathcal O(v^4/\mu_{TeV}^4)\right]
\label{RSpotential}
\end{equation}
can be approximated, as far as the phase transition from AdS-S to RS
is concerned, by the GW radion potential. In fact at the typical
radion transition temperatures the SM potential is a little
perturbation of the radion potential and should not alter the bounce
solution and the corresponding Euclidean action that governs the phase
transition. The radion potential then produces a supercooling of the
system and the electroweak phase transition takes place at much
lower temperatures than the typical EW ones and makes it to be much
stronger. We postpone the numerical analysis of this behaviour to
section~5.

In the approximation of Eq.~(\ref{totalpot}) the free energy is
minimized for $\mu=\mu_{TeV}$ and $\phi_c=v$, and at the minimum
\be
F_{c}^{min}(T)=-\,\frac{\pi^2}{90}g_c^*\,T^4\,.
\label{frsm}
\ee
Although irrelevant for what concerns the radion phase transition,
one-loop corrections to the Higgs potential are important at high
enough temperatures to determine the VEV of the Higgs field. For
$\mu=\mu_{TeV}$ the one-loop Higgs potential 
is just the standard one since
we have assumed a SM-like low energy dynamics, and is given in the
following.

In the $\overline{MS}$-scheme the one--loop Coleman-Weinberg potential
depends on the renormalization scale $Q$ through the ratio $m_i^2/Q^2$
while the thermal corrections similarly depend on the temperature
through the ratio $m_i^2/T^2$.  We can then write the SM potential at
$\mu=\mu_{TeV}$ as
\bea V_{SM}&=&
\frac{\lambda}{4}\left(\phi_c^2-v^2\right)^2
+\frac{1}{64\pi^2}\sum_{i=W,Z,h,\chi,t} n_i
[m_i(\phi_c)]^4\left[\log\frac{[m_i(\phi_c)]^2}{Q^2}-C_i\right]
\nn\\ &+& \frac{T^4}{2\pi^2}\left[\sum_{i=W,Z,h,\chi} n_i
J_B[m_i^2/T^2]+n_t J_F[m_t^2/T^2]
\right] ~,
\label{SMpotential}
\eea 
where the thermal fermionic (bosonic) function $J_{F(B)}$ are given by 
\begin{equation}
J_{F(B)}(y^2)=\int_0^{\infty} dx\
x^2\log\left[1 \pm e^{-\sqrt{x^2+y^2}}\right]~~,
\end{equation}
the constants $C_i$ derive from the $\overline{\rm MS}$
renormalization scheme we have used, and $n_i$ are the degrees of
freedom of the $i$ field, that is
\begin{eqnarray}
\label{csm}
C_W=C_Z  =  \frac{5}{6} ~~~~~~~~~
C_h=C_{\chi}=C_t  =  \frac{3}{2} \nonumber~ \\
n_W=6,\ n_Z=3,\ n_h=1,\ n_{\chi}=3,\ n_t=-12 ~.
\end{eqnarray}
The only fermion that contributes sizeably to the effective potential
for $\phi_c\neq 0$ is the top quark $t$ since the other fermions are
very weakly coupled to the Higgs doublet.  For practical purposes we
fix $Q=m_t(v)$ which is nearby the Higgs VEV.  As for the thermal
corrections it could have been possible to use a better approximation
including plasma effects~\cite{Dolan:1973qd},
however we think that for the purposes of the present paper the
one--loop approximation is good enough.

\section{The Phase Transition}

At high temperatures (early times) a deconfined (strongly--coupled) plasma
fills the Universe, which is described by the Friedmann--Robertson--Walker
metric $ds^2=dt^2-a^2(t)d\vec x^2$ and expands according to the Friedmann
equation
\be
\frac{\dot{a}}a\,=\,H\,=\sqrt{\frac{8\pi}{3 M_{P}^2}\rho_d}\,,
\label{feq}
\ee
where $\rho_d$ is the energy density of the plasma and can be easily
extracted from the free--energy in Eq.~(\ref{ultima}). As the Universe
expands, it cools down and a confining phase transition can occur when
the free--energy of the deconfined phase equals the confined one. This
happens at a critical temperature $T_c$ which is obtained by equating
Eqs.~(\ref{ultima}) and (\ref{frsm}):
\be
T_c^4\,=\,3\sqrt{(\theta\nu-\nu^2)|\epsilon|}
\left[\frac1{(1+\nu/2)^{3}}-\frac{2
\Delta g^*}{45 N^2}\right]^{-1}\left(\frac{\mu_{TeV}}\pi\right)^4\,.
\label{Tc}
\ee
where $\Delta g^*=g_c^*-g_d^*$ and $g^*_d$ is the number of light
degrees of freedom in the AdS-S phase that equals, in the model we are
considering, to the Higgsless SM degrees of freedom.  At small enough
$N$ and large enough $\nu$, $T_c^4$ can become negative. This is due
to the fact that the $T^4$ term in Eq.~(\ref{frsm}) becomes larger
than the one in Eq.~(\ref{ultima}) and in this case, given that $E_0$
is positive, the deconfined phase is never stable, the Universe stays
forever in the confined one and no phase transition occurs. However
for $N\geq 3$, and using $\Delta g^*=4$ in our model, this situation
is avoided provided that the constraints (\ref{perturbativity}) are
fulfilled \footnote{If some of the fermions, like for instance the top quark, 
are localized to the IR, $\Delta g^*$ would be larger. However, all our
results are substantially insensitive to its precise value.}.
In most of the allowed parameter space the critical
temperature is a bit smaller than $\mu_{TeV}$: $T_c\sim
\mu_{TeV}/\pi$.

\subsection{Completion of the transition}
Formally, the phase transition begins as soon as the temperature drops below
$T_c$. It will then be convenient, to fix ideas, to choose the origin of time
as $T(0)=T_c$ and rescale the space coordinates so that $a(0)=1$. There is a
simple one--to--one relation among temperature and time, $T(t)=T_c/a(t)$ and,
using Eq.~(\ref{feq})
\be
\dot{T}=-TH(T)=-\frac{T}{M_P}\sqrt{\frac{8\pi}3\rho_d(T)}\,.
\label{temptime}
\ee
For $T<T_c$ there is a non--zero probability for bubbles of the confined phase
to form, the rate of bubble nucleation for unit physical volume can be
expressed as
\be
\lambda(T)\,=\,A(T)\,e^{-S(T)}\,,
\label{nrate}
\ee
where the exponential suppression is due to semiclassical tunneling
and the bounce action $S(T)$ will be computed in the following. The
prefactor $A$ can be estimated on dimensional grounds, $A\sim T^a
T_c^b$ with $a+b=4$. In particular for very low temperatures, as those
that will be found in this work, $a=0,\, b=4$. The nucleation rate is
dominated by the exponential so that this rough estimate of the
prefactor will be sufficient for our purposes.

To have an idea of how fast the transition proceeds we
follow~\cite{Guth:1979bh} and write down the probability that one
point of space remains in the old (deconfined) phase at the time
$t$. Assuming the bubble expands at the speed of light
\bea
&&p=\exp\left[-\frac{4\pi}3\int_0^tdt_1a^3(t_1)
\lambda(T(t_1))\left(\int_{t_1}^tdt_2
\frac{1}{a(t_2)}\right)^3\right]\label{pdt}\\
&&=\exp\left[-\frac{4\pi}3\int_T^{T_c}dT_1
\frac{\lambda(T_1)}{T_1^4}\left(\frac{M_P}{\sqrt{8\pi/3}}\right)^4
\frac1{\sqrt{\rho_d(T_1)}}\left(\int_{T}^{T_1}dT_2
\frac1{\sqrt{\rho_d(T_2)}}\right)^3\right]\,,
\nn
\eea
where we used Eq.~(\ref{temptime}) to change variable in the time
integrals. The nucleation rate, due to the exponential factor in
Eq.~(\ref{nrate}), undergoes exponential variation in the course of
time, {\it i.e.} temperature, so that there could be regions of
temperature in which the $dT_1$ integral in Eq.~(\ref{pdt}) only
receives negligible contributions. Using dimensional analysis (we are
considering now $T\sim T_c\sim TeV$) to estimate $\rho_d$ and the
$dT_2$ integral in Eq.~(\ref{nrate}) (which we could as well compute
using Eq.~(\ref{ultima}) without changing the result which follows) we
find that the region of temperatures for which
\be
S(T)\,>\,B\,\equiv 4
\log\left(\frac{M_P}{\mu_{TeV}}\right)\approx 148
 \, \;\;\Rightarrow\;\;
\left[\frac{\lambda(T)}{T^4}\left(\frac{M_P}{T}\right)^4\ll1\;\;\;\;
\textrm{for}\;\;\;T\sim 1\;\textrm{TeV}\,\right],
\label{conds}
\ee
cannot contribute appreciably to the integral if $T$ is not orders of
magnitude smaller than $\mu_{TeV}\sim$ TeV.

We will soon see that, for $T\simeq T_c$, the condition (\ref{conds}) is
always fulfilled in the case we are considering, so that the phase transition
does not effectively start at $T=T_c$. As discussed above, it is very unlikely
to find a point in the new phase, the Universe is filled by a supercooled
deconfined plasma and keeps expanding with the energy density associated to
Eq.~(\ref{ultima}). It is worth remarking that, as the temperature drops below
\be
T_i^4\,=\,\sqrt{(\theta\nu-\nu^2)|\epsilon|}(1+\nu/2)^{3}
\left(\frac{\mu_{TeV}}\pi\right)^4\,<\,T_c^4/3\,,
\ee
the cosmological constant starts to dominate over radiation and an
inflationary epoch begins. The total number of e--folds of inflation
is $\sim \log T_i/T_n$ where $T_n$, to be defined below, is the
temperature at which the transition effectively occurs. In the
inflationary epoch we can use $\rho_d=E_0$ in Eq.~(\ref{pdt}) and
write down more explicitly the probability of transition as
\be
p(T)=\exp\left[-\frac{4\pi}3\int_T^{T_i}\frac{dT_1}{T_1}
\frac{\lambda(T_1)}{E_0^2}\left(\frac{M_P}{\sqrt{8\pi/3}}\right)^4
\left(1-\frac{T}{T_1}\right)^3\right]\,,
\label{pr}
\ee
where we neglected the $[T_i,T_c]$ part of the integral. Once again we
can check that temperatures for which the condition (\ref{conds})
holds do not contribute to the integral. In our model, the bounce
action $S(T)$ monotonically decreases with temperature so that if
\textbf{(a)} $S(T\rightarrow0)>B$ the action never crosses the
critical value while if \textbf{(b)} $S(T\rightarrow0)<B$ we can
define the nucleation temperature $T_n$ as
\be
S(T_n)\,=\,B\,.
\label{150}
\ee
The case \textbf{(a)} corresponds to the ``slow'' phase transition
considered in~\cite{Guth:1982pn} and, despite of the fact that the
probability (\ref{pr}) will eventually go to zero, bubbles will not
percolate, the phase transition will never end and we will have an
empty and cold Universe. Avoiding \textbf{(a)} will give us a strong
constraint on the allowed parameter space of the model and, in
particular, on the maximum value of the number of colors $N$.

Let us now focus on case \textbf{(b)}. For $T_1=T_i$ the nucleation
rate is always too small for the transition to start but for $T_1=T_n$
the argument of the integral in Eq.~(\ref{pr}) becomes of order
one. It is important to remark that the bounce action has a
power--like behaviour with temperature so that the rate grows
exponentially. When the temperature crosses $T_n$ the argument of the
integral suddenly passes from being much smaller to much bigger than
one. The integral suddenly becomes large, the probability goes to zero
and the phase transition ends. During the time the phase transition
lasts, therefore, the Universe undergoes a negligible cooling, {\it
i.e.} a negligible expansion. On the scale of the evolution of the
Universe the phase transition is instantaneous: bubbles expand,
collide and percolate as if the Universe were static.

\begin{figure}[htb]
\psfrag{x}[][bl]{$T/T_n$}\psfrag{p}[][l]{$p(T)$}
\psfrag{y}[][bl]{$\log_{10}(T/T_i)$}
%
\vspace{.5cm}
\begin{center}
\epsfig{file=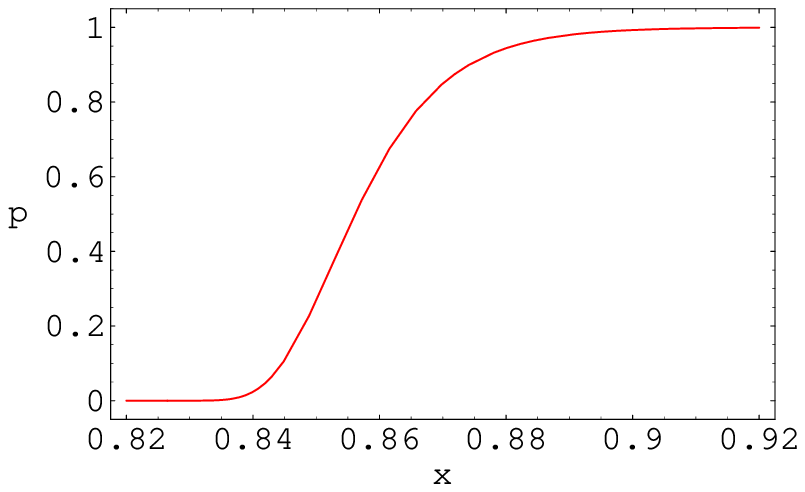,width=0.45\textwidth}\hspace{.5cm}
\epsfig{file=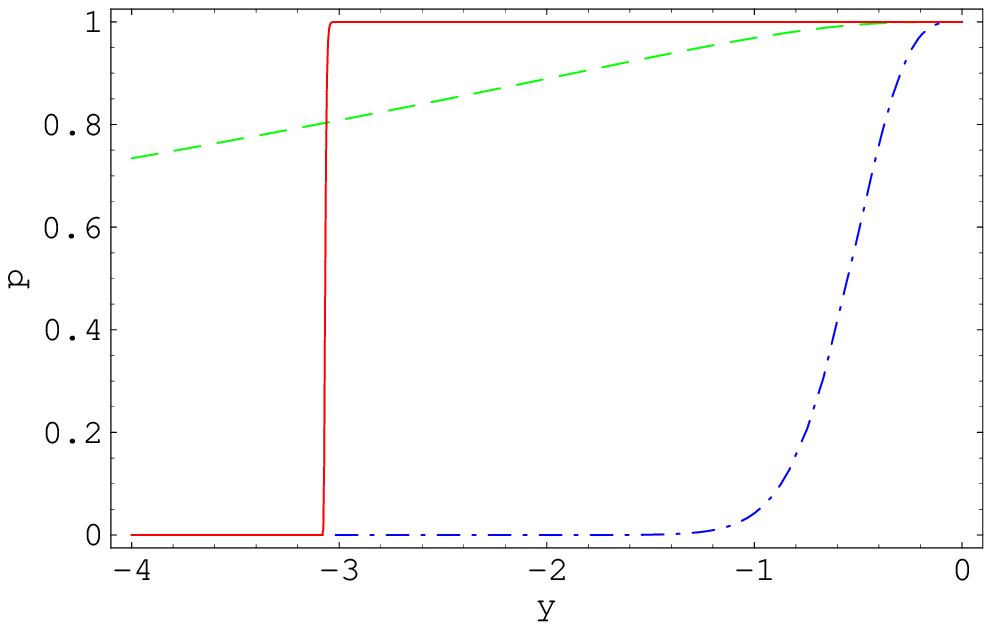,width=0.45\textwidth}
\end{center}
\caption{Left panel: Plot of the probability (\ref{probinf}) as a
  function of $T/T_n$ for $T_n=10^{-3} T_i$ and a slope
  $\beta=100$. Right panel: Comparison of the previous plot [red
  (solid) line] as a function of $\log_{10} T/T_i$ with the
  probability (\ref{probguth}) for $\epsilon_0=1$ [blue (dot-dashed)
  line] and $\epsilon_0=.01$ [green (dashed) line].}
\label{probabilidad}
\end{figure}

We can illustrate this behaviour with a simple toy model. Since the
integral (\ref{pr}) is dominated by temperatures very close to $T_n$
one can assume that the action in (\ref{nrate}) behaves linearly with
$T$, i.e.~$S(T/T_n)=\alpha+\beta\, T/T_n$ where $\alpha+\beta\sim B$
is the condition (\ref{150}) and the slope $\beta$ has been computed
numerically and yields values $\beta\sim 50-100$ (depending on the
values of the GW parameters). We can then write for the probability
(\ref{pr}) the expression
\be 
p(T)=\exp\left[-\frac{4\pi}{3}\int_{T}^{T_i}\frac{dT_1}{T_1} e^{B} 
e^{-(\alpha+\beta\, T_1/T_n)}\left(1-\frac{T}{T_1}\right)^3\right]\;.
\label{probinf}
\ee
The behaviour of the probability $p(T)$ is shown
in the left panel of Fig.~\ref{probabilidad}.
We see that, in this particular case the phase transition happens
between $T=0.88\,T_n$ and $T=0.84\, T_n$, a negligible cooling that
corresponds to the little expansion of 0.05 e-folds during the phase
transition. The model considered in Ref.~\cite{Guth:1982pn}
corresponds to the case where $\beta=0$ and Eq.~(\ref{probinf}) reads
\be
p(T)=\exp\left[-\frac{4\pi}{3} \epsilon_0 \left(N_e-\frac{11}{6}
+3e^{-N_e}-\frac{3}{2}e^{-2N_e}+\frac{1}{3}e^{-3N_e}\right) \right]
\label{probguth}
\ee
where $\epsilon_0=\exp(B-\alpha)$ and we can see that a large number of
e-folds $N_e(T)=\log(T_i/T)$ is related to the value of $\epsilon_0$ by
$N_e\sim 3/4\pi\epsilon_0$ which would imply $\epsilon_0\ll 1$. On the other
hand for $\epsilon_0\sim 1$ we obtain $p\ll 1$ after an $\mathcal O(1)$ number
of e-folds of inflation. This does not happen in our case since nucleation is
triggered by the temperature $T_n$ where condition (\ref{150}) holds. The
difference between both situations is illustrated in the right panel of
Fig.~\ref{probabilidad}.

An analogous situation was discussed in Ref.~\cite{Turner:1992tz} and
it was shown that two
conditions must be satisfied by $\gamma\equiv\lambda/H^4$ to complete
the phase transition in an (inflationary) first-order phase
transition. {\bf i)} $\gamma>9/4\pi$ is required for percolation; {\bf
ii)} $\gamma\simeq 0$ for almost all the inflationary period,
otherwise bubbles are formed, grow during inflation and distort the
CMB radiation. We have checked that in our model both conditions are
satisfied at least for moderate number of e-folds of inflation, {\it
i.e.} for $N_e\leq 6$. Certainly condition {\bf i)} is satisfied since
$\gamma$ crosses from values larger to values smaller than one around
the nucleation temperature. Condition {\bf ii)} is also satisfied
since $H$ is a constant during the inflationary period while $\Gamma$
varies exponentially with the temperature so that $\gamma$ varies very
quickly near the nucleation temperature. Therefore big bubbles do not
have the chance to develop. We did not make the numerical analysis for
the very small values of the nucleation temperature corresponding to
the number of e-folds of inflation consistent with the required amount
of cosmological inflation $(N_e\sim 30)$ because inflation was not the
main issue in this paper. We will come back to this issue in a
different work.

Summarizing, we have seen that, provided Eq.~(\ref{150}) is verified 
at some non--zero temperature, the phase transition completes,
the new phase plasma thermalizes and the expansion continues with the
free--energy given by Eq.~(\ref{frs1}). During the transition, of course, the
energy (or what is the same, given that the Universe is effectively static,
 the energy density) must be conserved so that the Universe
will end up in the confined
phase with a reheating temperature $T_R$ given by
\begin{equation}
\frac{\pi^2}{30} g^*_c\,T_R^4=E_0\,.
\label{TR1}
\end{equation}
To obtain the above equation we equated $\rho_c(T_R)$ obtained from
Eq.~(\ref{frs1}) with $\rho_d(T_n)$ and neglected the nucleation
temperature $T_n\ll T_i$.

\subsection{Bubble nucleation: approximate analytical results}

We will now describe in detail the tunneling process from the deconfined
(AdS-S) to the confined (RS) phase as described by the respective
free-energies given by Eqs.~(\ref{fadss1}) and (\ref{frs1}).  At the critical
temperature $T_c$ the two minima are degenerate and thus the Euclidean action
$S(T_c)\to\infty$. Below $T_c$ the first bubbles that can be formed are those
with $O(3)$ symmetry, thin walls and very large radii. In this case the
probability of bubble formation is governed by $S=S_3(T)/T$ where $S_3$ is the
Euclidean action with $O(3)$ symmetry. In the thin wall approximation the
action can be given an analytic expression as
$
S_3=16 \pi S_1^3/3 (\Delta V)^2
$
where $S_1$ is the surface tension that can be evaluated in the limit $T\to
T_c$ and $\Delta V$ is the depth of the true vacuum relative to the false
one~\cite{Coleman:1977py,Linde:1981zj}.
Using the analytic approximation we will prove now that the radion phase
transition will not allow the formation of thin wall bubbles. For that we will
momentarily forget the Higgs field since its presence will not alter the
following conclusions. A straightforward application of the thin wall formula
for the deconfinement/confinement radion phase transition leads to
\be 
\frac{S_3(T)}{T}\simeq \frac{64\pi}{3}
\frac{\mu_{TeV}^2}{\sqrt{2E_0}}
\left(\frac{3
N^2}{2\pi^2}\right)^{3/2}
{\displaystyle
\frac{\mu_{TeV}/T}
{\left(1-T^4/T_c^4\right)^2} }   ~,
\label{S3T}
\ee
where $E_0$ is the vacuum energy defined in (\ref{vacuumE}). Eq.~(\ref{S3T})
shows that $S_3(T)/T$ goes to infinity in both limits $T\to 0$ and $T\to T_c$.
The function $S_3(T)/T$ has in fact a minimum (that corresponds to the maximal
probability of nucleation) at $T^*=T_c/\sqrt{3}$.  Using now the relation
(\ref{Tc}) between $\mu_{TeV}$ and $T_c$ one can write the final expression
for the action (\ref{S3T}) at $T=T^*$ as
\be
\frac{S_3(T^*)}{T^*}\simeq\frac{50 N^2}{[(\theta-\nu)\nu|\epsilon|]^{3/8}}
\left[\frac{1}{(1+\nu/2)^3}-\frac{2 \Delta g^*}{45 N^2}\right]^{1/4}
>\frac{50 N^2}{|\epsilon|^{3/8}} ~,
\label{final}
\ee
where the last inequality follows from the fact that the term raised to the
power $1/4$ is an order one number and from the perturbativity bounds
$\theta<1,\ \nu<1$. For instance for $|\epsilon|\sim 1/10$, $S_3/T$ does not
satisfy condition (\ref{150}) for $N\geq 2$ 
while $N \gtrsim 3$ is needed for a reliable $1/N$ (semiclassical)
 expansion of our theory.
We can observe that if the values
of the parameters $(\theta,\nu)$ are small, as it corresponds to the
perturbativity region, the values of the Euclidean action $S_3/T$ can be huge.

Of course thin wall bubbles with symmetry $O(4)$ have Euclidean actions larger
than those just described with $O(3)$ symmetry. On the other hand since there
is supercooling we will found that in most of the available parameter space
the least action corresponds to $O(4)$ symmetric solutions.

For $O(4)$ symmetric bubbles we have worked out the thick wall approximation
that we will illustrate now and where we have also included the Higgs
potential. In order to do that we will work out the thick wall approximation
for a set of field with (in general) non-canonical kinetic terms as follows.
Let us consider a theory with a number of scalar fields
$\phi\equiv(\phi_1,\phi_2,\dots)$ with a potential $V(\phi)$ and general
non-canonical kinetic functions $f_i(\phi)$. The euclidean action for $O(4)$
configurations can be written as
\begin{equation}
S_4=2\pi^2\int_0^\infty d\rho \rho^3\left(\sum_i
  f_i(\phi)\frac{1}{2}\phi^{\prime\, 2}_i+V(\phi)\right)~,
\label{action4}
\end{equation}
where we are using the notation $\phi^\prime=d\phi/d\rho$.  The potential $V$
has a false vacuum (say at $\tilde\phi=0$) where the energy is normalized to
be zero $V(\tilde\phi)=0$ and a true minimum at $\phi=\phi_0$. The bounce
solution for $O(4)$ symmetric bubbles is obtained by solving the equations of
motion (EOM)
\begin{equation}
  f_i\phi_i^{\prime\prime}+
\frac{3}{\rho}f_i\phi^\prime_i+\phi^\prime_j\frac{df_i}{d\phi_j}\phi^\prime_i
=\frac{dV}{d\phi_i}+\frac{1}{2}\frac{d f_j}{d\phi_j} \phi^{\prime\, 2}_j
  \quad \textrm{no summation on i} ~,
\label{eom}
\end{equation}
with the boundary conditions $\phi^\prime(\rho=0)=0$ and
$\phi(\rho\to\infty)=\tilde\phi$.

We can now consider a bubble of true vacuum and radius $R$ inside an
exterior of false vacuum. Since the potential outside the bubble is
zero and the $\phi$-profile constant, only the region inside the
bubble will contribute to the action (\ref{action4}). If
$\phi^*\equiv\phi(0)$ is the value of $\phi$ at which the bounce
starts~\footnote{Notice that $\phi^*$ does not coincide with
$\phi_0$.}, $\delta\phi=\phi^*-\widetilde\phi$ is the total variation
of $\phi$ inside the bubble wall $\delta R$~\footnote{The bubble wall
$\delta R$ is defined as the region $\rho$ where $\phi$ varies.}. We
can then approximate the action (\ref{action4}) by
\begin{equation}
S_4\simeq \pi^2 R^3 \sum_i f_i \left(\frac{\delta \phi_i}{\delta R}\right)^2
\delta R+\frac{\pi^2}{2}\overline V R^4 ~,
\label{actionapp}
\end{equation}
where $\ov V<0$ is a suitable average of the potential inside the bubble. We
will take $\ov V=V(\phi^*)$ and will use the notation $V(\phi^*)\equiv V^*$
and so on. Notice that in this way we will underestimate the bounce action.

If the bubble is thick we take $\delta R=R$ and determine the radius $R_c$ of
the bubble by extremizing Eq.~(\ref{actionapp}) with respect to $R$. This
gives for the critical radius
\begin{equation}
\label{critR}
R_c^2=\frac{\sum_i\phi_i^{*\, 2}f_i^*}{-V^*}~,
\end{equation}
and for the critical action
\begin{equation}
\label{actioncrit}
S_4(R_c)=\frac{\pi^2}{2}\frac{\left(\sum_i \phi_i^{*\, 2}f_i^*\right)^2}{-V^*}~.
\end{equation}

We will apply the previous expressions to the case of the system of two
fields, radion and Higgs fields, $(\mu,\phi_c)$ with non-canonical kinetic
functions
\begin{equation}
\mathcal L_{kin}=\frac{3N^2}{2\pi^2}\frac{1}{2}
(\partial_\nu\mu)^2+\frac{\mu^2}{\mu_{TeV}^2} |\partial_\nu h|^2~,
\end{equation}
and potential 
\begin{equation}
V\equiv \mu^4\left(\mathcal V_{GW}+\frac{\mathcal
V_{SM}}{\mu_{TeV}^4} \right)~,
\label{V}
\end{equation}
where~\footnote{For the purpose of this section it is enough to just
consider the tree level SM potential with radiatively corrected
parameters.}
\begin{eqnarray}
\mathcal
V_{GW}(\mu)&=&(4+2\epsilon)\left[v_1-v_0
\left(\frac{\mu}{k}\right)^\epsilon\right]^2+\delta
  T_1-\epsilon v_1^2 \nonumber\\
\mathcal V_{SM}(\phi_c)&=&\frac{1}{4}\lambda(\phi_c^2-v^2)^2 ~.
\label{potenciales}
\end{eqnarray}
The value of the critical action is then given by
\begin{equation}
S_4=\frac{\pi^2}{2}\, \frac{\left({\displaystyle
\frac{3N^2}{2\pi^2}+\frac{\phi_c^{*\, 2}}{\mu_{TeV}^2
}}\right)^2}{-\mathcal V_{GW}^{\,*}-\mathcal V_{SM}^{\,*}}~,
\label{accion}
\end{equation}
which depends on the values of $\mu^*$ and $\phi_c^*$ that in turn do not
necessarily coincide with their vacuum values $\mu_{TeV}$ and $v$
respectively.  In fact the precise values of $\mu^*=\mu(0)$ and
$\phi_c^*=\phi_c(0)$ require of a numerical calculation of the bounce
solution. Here we just want to give a plausibility argument for the phase
transition to happen.

The phase transition should happen when $S_4\sim B$ as we have seen in
(\ref{150}).  The actual value of $S_4$ in Eq.~(\ref{accion}) depends, for
fixed values of the GW potential parameters, on the fields $\mu^*$ and
$\phi_c^*$. We can minimize the action $S_4$ with respect to those fields in
order to obtain the most favorable configuration for the bounce solution.
Minimization with respect to $\mu^*$ yields the value
\begin{equation}
\mu^*=k\left(\frac{v_1}{v_0}\right)^{1/\epsilon} ~,
\end{equation}
for which the thick-wall action is found to be
\begin{equation}
\label{accion2}
S_4(\phi_c^*)=\frac{\pi^2}{2}\, \frac{\left({\displaystyle
\frac{3N^2}{2\pi^2}+\frac{\phi_c^{*\, 2}}{\mu_{TeV}^2
}}\right)^2}{\epsilon v_1^2-\delta T_1-\mathcal V_{SM}^{\,*}} ~.
\end{equation}
The action (\ref{accion2}) has a minimum at $\phi_c^*=0$ for light
Higgs masses. However the point $(\mu^*,0)$ is a saddle point of the
potential and the corresponding configuration is classically
stable. This situation is avoided for values of $\phi_c^*\neq
0$. Since the function $S_4(\phi_c^*)$ is monotonically increasing
between $\phi_c^*=0$ and $\phi_c^*=v$ a conservative assumption would
be to fix $S_4(v)<B$ and therefore this condition will be satisfied by
all configurations with $0<\phi_c^*\leq v$.  The $S_4(v)<B$ condition
can be expressed as
\begin{equation}
\theta-\nu>\frac{3 N^2}{2 B}
\left(1+\frac{2 \pi^2}{3 N^2}\frac{v^2}{\mu_{TeV}^2}\right)^2\simeq
\frac{3 N^2}{2 B}\left(1+\frac{0.4}{N^2}\right)^2 ~,
\label{thickbound}
\end{equation}
where the second term inside the squared is a very tiny correction
reflecting the fact that the SM potential is a small perturbation of
the GW potential. 

\subsection{Bubble nucleation: numerical results}

Generally speaking, the approximation methods used in the previous
section are barely reliable. The main limitation is that they do not
provide a reliable determination of the nucleation temperature as it
is not possible to know, a priori, which shape (thick or thin) the
bounce solution has at a given temperature. A numerical calculation of
the bounce is therefore needed, and in the case of a single scalar
field the solution can be easily found iteratively by the technique of
over--shoots and under--shoots~\footnote{We have explicitly checked
our bounce program for the potentials of Ref.~\cite{Linde:1981zj}}.
Clearly this can be done at zero and finite temperature.

In our case we have three fields ($T_h$, $\mu$ and $h$) involved in
the bounce but, as we have discussed before, the Higgs does contribute
to the bubble action with a tiny amount. Moreover the two remaining
fields follow a single path in the two-dimensional field configuration
space~\cite{Creminelli:2001th} and then we can consider the transition
for a single field defined as
\begin{eqnarray}
\Phi \equiv \left\{\begin{array}{l} 
          - \left(N \sqrt{3/2}/\pi \right)
          T_h  \hspace{2cm} \textrm{for}\; \Phi < 0  \\
          \left(N \sqrt{3/2}/\pi \right) \mu
	  \hspace{2.6cm} \textrm{for } \;\Phi >0  \; ,

          \end{array}\right.         
\end{eqnarray}
which has a canonical kinetic term~\footnote{We are supposing for
  $T_h$ the kinetic term $\frac{3N^2}{2\pi^2}\frac{1}{2}(\partial_\nu
  T_h)^2$ similar to the radion one.} and the potential~\footnote{This
  potential is a simplification of that considered in the numerical
  analysis. Since for $T\neq 0$ the potential is discontinuous at
  $\Phi=0$, it has been regularized and checked that the result is not
  sensitive to the regularization.}
\bea
V(\Phi,T)= F_d \left(\pi \sqrt{2/3}/N\, \Phi,T\right) \, \Theta(-\Phi) +
           F_c \left(\pi \sqrt{2/3}/N\, \Phi,0,T\right) \, \Theta(\Phi)~,
\eea
where $\Theta$ is the Heaviside step function.

\begin{figure}[]
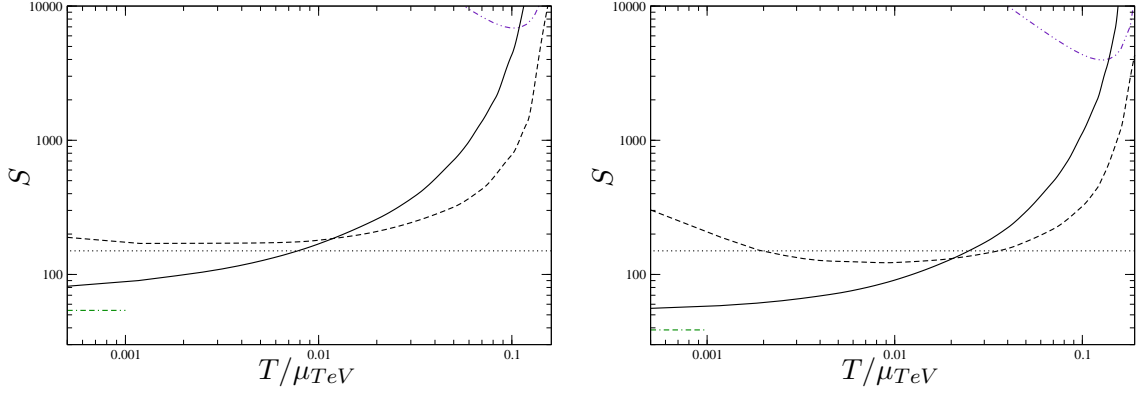

\vspace{0cm}
\begin{center}
\psfrag{t}[][bl]{$T/\mu_{TeV}$}\psfrag{f}[][l]{$S$}
\epsfig{file=actionBubbles.eps,width=6.5cm,bbllx=85,
bblly=-44,bburx=703,bbury=390}
\hspace{1cm}
\epsfig{file=actionBubbles2.eps,width=6.5cm,bbllx=85,
bblly=-44,bburx=703,bbury=390}
\vspace{-1.7cm}
\end{center}
\caption{Solid [dashed] lines is the plot of the $O(4)$ [$O(3)$]
Euclidean action as a function of $T/\mu_{TeV}$ for $N=3$, $\epsilon =
-1/20$ and, for the left [right] panel, $(\theta,\nu)=(0.35,0.1)$
[$(\theta,\nu)=(0.6,0.25)$]. On the left [right] the transition occurs
via $O(4)$ [$O(3)$] bubbles. The analytical thin wall $O(3)$
(\ref{final}) is plotted in double-dotted-dashed and the $O(4)$
analytical asymptotic value (\ref{accion2}) in double-dashed-dotted.}
\label{bubbles}
\end{figure}

Figure~\ref{bubbles} provides an example of the numerical results we
 find for the bounce action [$S=S_4$ or $S=S_3/T$] with $O(3)$ and
 $O(4)$ symmetry in two typical points of our parameter space of
 $N=3$.  The comparison with analytical (thick and thin) formulas
 shows up a considerable discrepancy. The nucleation rate, at a given
 temperature, is dominated by the bounce of minimal action and,
 depending on the point in the parameter space, bounces 
\begin{figure}[]
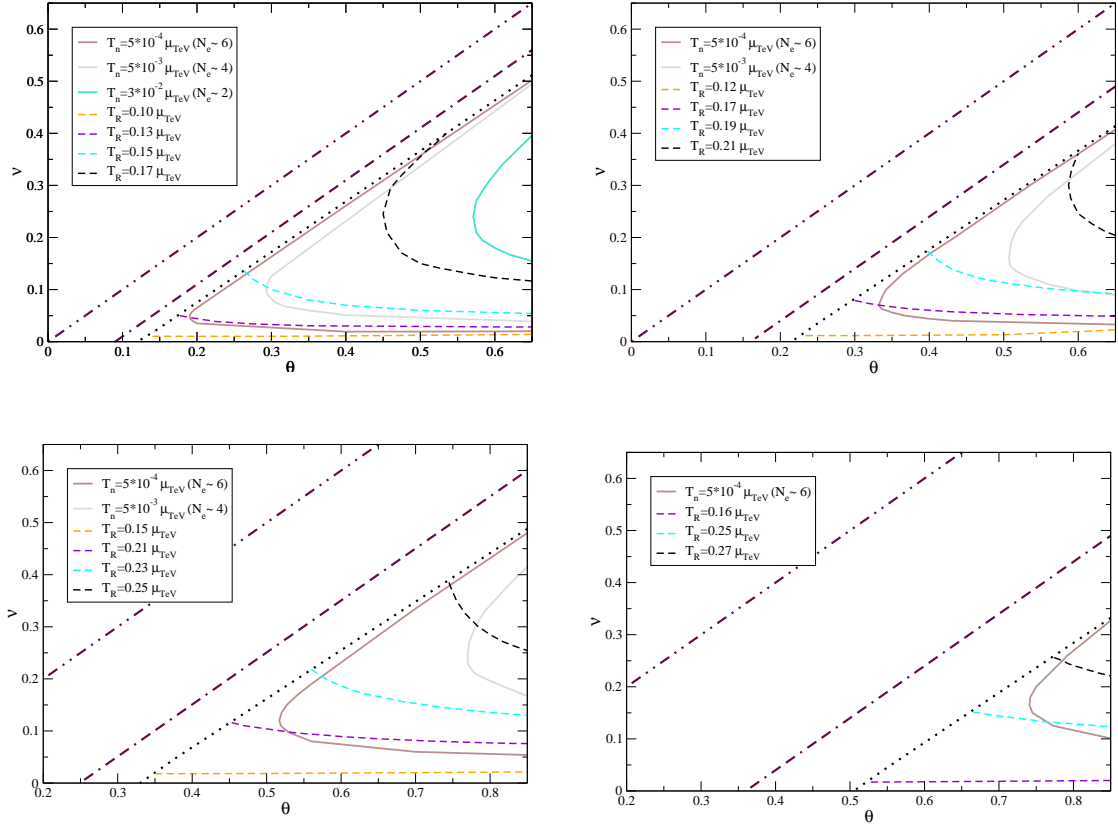

\begin{center}
\vspace{1.6cm}
\epsfig{file=N3.eps,width=6.5cm,bbllx=85,bblly=-34,bburx=703,bbury=390}
\hspace{1cm}
\epsfig{file=N4.eps,width=6.5cm,bbllx=85,bblly=-34,bburx=703,bbury=390}
\epsfig{file=N5.eps,width=6.5cm,bbllx=85,bblly=76,bburx=703,bbury=521}
\hspace{1cm}
\epsfig{file=N6.eps,width=6.5cm,bbllx=85,bblly=86,bburx=703,bbury=521}
\vspace{-.1cm}
\end{center}
\caption{Plot of the allowed regions in the plane $(\theta,\nu)$ for
$N=3$ (upper-left panel), $N=4$ (upper-right panel), $N=5$ (lower-left
panel) and $N=6$ (lower-right panel), with $\epsilon=-1/20$ and $v_0$
as a function of $\mu_{TeV}$. The consistency bound $\theta<\nu$ is
marked by a double-dotted-dashed line.  The region where the
nucleation is allowed is on the right of the dotted line, obtained by
numerical $O(4)$ bubbles at $T_n=0$. The same upper bound, but
calculated using analytical approximations, is the
double-dashed-dotted line.  The points where the nucleation occurs at
some fixed temperature are indicated by colored solid lines.  Finally,
several reheating temperature curves are drawn in dashed.}
\label{bounce}
\end{figure}
with either
 symmetries can be dominant. In all cases, however, the nucleation
 temperature is quite small. For the cases with a moderate number of
 e-folds ($N_e\leq 6$) that we have analyzed numerically we have
 checked that the slope of the Euclidean action $S(T/T_n)$ is
 $\beta\sim 50-100$ as it was anticipated in section~4.1.

Most of our numerical results are summarized in Fig.~\ref{bounce}
 which shows, for $\epsilon=-1/20$ and different values of $N$, the
 triangle in the $\theta$-$\nu$ plane allowed by consistency
 (\ref{perturbativity}) and, inside that, the analytic bound for the
 nucleation to happen as obtained by Eq.~(\ref{thickbound}). Contour
 plots are shown for fixed nucleation temperature $T_n=5\cdot
 10^{-4}\mu_{TeV}$, $T_n=5\cdot 10^{-3}\mu_{TeV}$ and $T_n=5 \cdot
 10^{-2} \mu_{TeV}$, corresponding to an inflation of $N_e\sim 6,4,2$,
 respectively. Regions of higher nucleation temperature, {\it{i.e.}}
 smaller inflation, are to the right of such level curves. As we can
 see in the allowed region a supercooling is necessary before
 permitting nucleation, and the larger $N$ the lower $T_n$ as
 predicted by the analytical approximations.  In the allowed area,
 $O(4)$ bubbles dominate over $O(3)$, except in the case $N=3$ when
 $T_n \gtrsim 10^{-2} \mu_{TeV}$.  At $T=0$ the GW potential, for
 $\epsilon<0$, has a local minimum at $\mu=0$ and zero--temperature
 nucleation occurs as a tunneling from this secondary minimum to the
 global one. By studying this zero--temperature tunneling\footnote{See
\cite{Cline:2000xn} for an early study of this transition.}
 we have obtained numerically the region in which nucleation
 is allowed (dotted line in the Fig.~\ref{bounce}). We see
 that the analytical formula provides a reasonable approximation.
 Also in Fig.~\ref{bounce} level contours for fixed values of the
 reheating temperature in units of $\mu_{TeV}$ (dashed lines) are
 shown.

\begin{figure}[h]
\vspace{.7cm}
\begin{center}
\epsfig{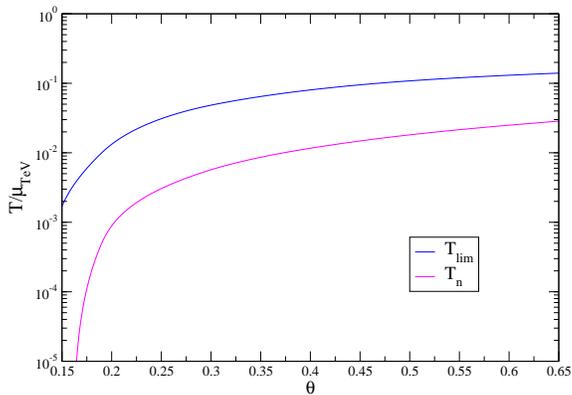}
\end{center}
\caption{Lower curve: Plot of $T_n/\mu_{TeV}$ as a function of
$\theta$ along the bisectrix [$\nu =\theta/2-1/20$] of the allowed
region in the upper-left plot of Fig.~\ref{bounce} $(N=3)$. Upper
curve: Similar plot for $T_{lim}\equiv 1/(2 R) \mu_{TeV}$ where $R$ is
the radius of the $O(4)$ bubble in the thick wall approximation.}
\label{nucleation}
\end{figure}

Until now we have considered $O(4)$ solutions also at finite
temperature, but strictly speaking they have sense only if the
diameter of the bubble $2 R$ is smaller than $T_n^{-1}$, the size of
the compactified dimension at the nucleation temperature. Under that
circumstance $O(4)$ is a good approximate
symmetry~\cite{Linde:1981zj}.
To show that explicitly we plot in
Fig.~\ref{nucleation} the nucleation temperature $T_n$ (without
considering $O(3)$ solutions) and the corresponding $T_{lim}\equiv
1/(2 R) \mu_{TeV}$ as a function of $\theta$ along the bisectrix 
[$\nu=\theta/2-1/20$] of the allowed
region of Fig.~\ref{bounce} for
$N=3$~\footnote{R is calculated using the $O(4)$ thick wall
approximation.}. The restriction of this analysis to the bisectrix in
$N=3$ is justified by the fact that along this path the maximal
nucleation temperatures are reached and therefore it should be the
worse region for the $O(4)$ approximate symmetry. From the
plot we can conclude aposteriori that $O(4)$ actions considered in
Fig.~\ref{bounce} make sense.

\section{Conditions for electroweak baryogenesis}

As we have seen the electroweak phase transition in the RS model,
triggered by the radion phase transition, is a strong (supercooled)
first-order one. This is unlike the pure Standard Model where the
electroweak phase transition is very weakly first order at any
perturbative level~\footnote{In fact non-perturbative analyses have
shown that it is so weak that it disappears (it becomes a continuous
crossover) when non-perturbative effects are taken into account.}.
Since the weakness of the electroweak phase transition in the Standard
Model was one of the main obstacles to the mechanism of electroweak
baryogenesis we can see that the situation is completely different in
the Randall-Sundrum setup.  

The first condition to be imposed for a successful EWBG is the 
non-erasure of the generated
baryon asymmetry at the nucleation temperature inside the bubbles. As
we can easily check this does not really impose any additional
condition on the model parameters. In fact because the phase
transition is strongly first order, and in particular
$\langle\phi_c(T_n)\rangle/T_n\gg 1$, sphalerons are totally inactive
inside the bubbles and the baryon asymmetry generated by some extra
source of $CP$-violation (on top of the standard CKM phase) will not
be erased in the broken phase.

The second condition is that sphalerons have to be active outside the
bubbles and in order to check this we need an estimate of the
sphalerons rate in the symmetric (deconfined) phase. The SM gauge
bosons and the fermions arise from bulk fields in our scenario, so
that they correspond holographically to fundamental $4D$ fields which
are mixed with the strongly coupled $CFT$. In the deconfined phase,
then, we just have the SM fields (except for the Higgs) mixed with the
new sector.  Neglecting the mixing, which is expected to give a small
New Physics correction, the sphaleron transitions are just like in the
Higgsless SM. In the symmetric phase (see
{\it{e.g.}}~\cite{Ambjorn:1988gf}) the presence of the Higgs is
irrelevant for the calculation of the sphaleron rate and we can then
simply use the SM results. On dimensional
grounds~\cite{Arnold:1987mh}, the sphaleron rate can be estimated as
$\Gamma_{sph}\sim \kappa \alpha_W^4 T^4$ where the coefficient
$\kappa$ has been evaluated numerically in~\cite{Ambjorn:1988gf} with
the result $0.1\leq\kappa\leq1$ and $\alpha_W$ is the weak gauge
coupling constant. The rate of baryon number non-conserving processes
$V_B(T)$ is related to $\Gamma_{sph}$ by~\cite{Shaposhnikov:1987tw}
\be
V_B(T)=\frac{13}{2} N_f\frac{\Gamma_{sph}}{T^3} ~,
\label{VB}
\ee
where $N_f=3$ is the number of generations. We have to compare now the
rate (\ref{VB}) with the expansion rate of the Universe given by
\be 
\chi=\left(\frac{2
N^2}{\pi}\right)^{1/2}\left[(\theta\nu-\nu^2)|\epsilon|\right]^{1/4}
\frac{\mu_{TeV}^2}{M_{Pl}} ~.
\ee
This gives a lower bound on the nucleation temperature provided by
\be
\frac{T_n}{\mu_{TeV}}> 3\times 10^{-12} N 
\left[(\theta\nu-\nu^2)|\epsilon|\right]^{1/4}  ~,
\label{boundTn}
\ee
which corresponds to an upper bound on the number of e-folds of
inflation as $N_e<26$. The
nucleation temperature widely satisfies the
bound (\ref{boundTn}) in the region we have analyzed.  

As we have seen the electroweak phase transition in our model is
accompanied by a period of exponential expansion corresponding to a
(few) number of e-folds of inflation. After the phase transition the
Universe is reheated to a given temperature $T_R$. If the temperature
$T_R$ is higher than the temperature at which $\phi_c(T)/T=1$
sphalerons will be activated inside the bubbles and the generated
baryon asymmetry will be erased. So a third condition is that the
reheating temperature be lower than the temperature at which
$\phi_c(T)/T=1$ for a fixed value of the Higgs mass.

\begin{figure}[htb]
\vspace{.4cm}
\begin{center}
\psfrag{t}[][bl]{\footnotesize{$ T [GeV]$}}
\psfrag{n}[][l]{\footnotesize{$\langle\phi_c(T)\rangle / T$}}
\psfrag{Higgs[GeV]}[][]{\footnotesize{$m_H [GeV]$}}
\psfrag{Treh[GeV]}[][]{\footnotesize{$T_R^{max}[GeV]$}}
\epsfig{file=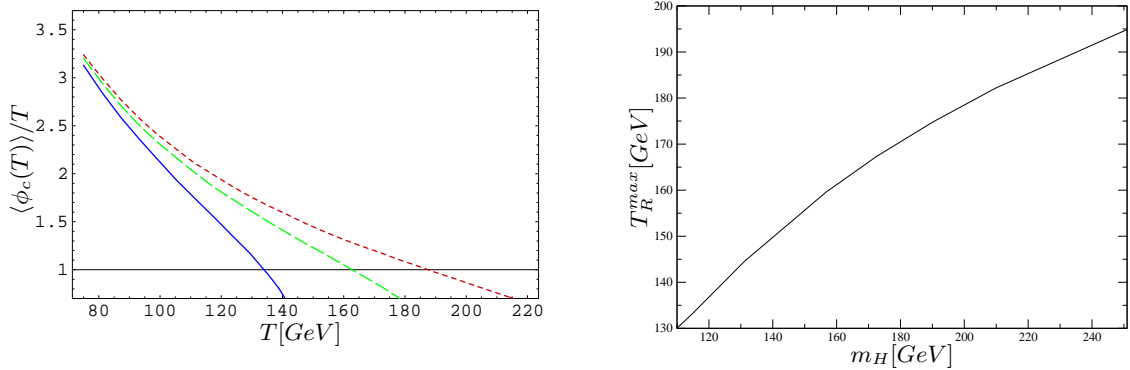,width=7cm,bbllx=90,bblly=6,bburx=375,bbury=182}
\hspace{1cm}
\epsfig{file=maxTreh.eps,width=6.5cm,bbllx=37,bblly=71,bburx=703,bbury=536}
\end{center}
\caption{Left panel: Plot of $\langle\phi_c(T)\rangle / T $ as a
function of the temperature in GeV for a Higgs mass of 115 GeV (solid
line), 165 GeV (dashed) and 225 GeV (short dashed). Right panel: Plot
of the temperature at which the SM potential is minimized at
$\langle\phi_c(T)\rangle/T=1$ as a function of the Higgs mass in
GeV.}
\label{SM-V_T}
\end{figure}

In order to understand the temperature at which $\phi_c(T)/T=1$ for
the potential (\ref{SMpotential}), we plot the
value of the ratio $\langle\phi_c(T)\rangle/T$ versus the temperature
$T$ (Fig.~\ref{SM-V_T}).  Here the three lines are calculated for a
Higgs of 115 GeV (solid line), 165 GeV (dashed) and 225 GeV (short
dashed).  In correspondence to these masses the critical temperature
$T_c^{SM}$~\footnote{Here we define the critical temperature
$T_c^{SM}$ as the temperature where the non-trivial minimum of
$V_{SM}$ and the origin are degenerate.} has been found, resulting
respectively 146, 195 and 250 GeV, at which
$\langle\phi_c(T_c)\rangle/T_c$ results approximately 0.2, 0.1 and
0.06~\footnote{Of course for such small values of $\phi_c$ our
approximation fails as a consequence of the IR problem of thermal
field theories.}.  A remarkable feature of the SM potential is that,
inside a temperature interval of 1 GeV, the origin passes from being a
minimum to a maximum and during this change $\langle\phi_c(T)\rangle$
runs incredibly fast.  Clearly this behaviour of
$\langle\phi_c(T)\rangle$ remains also at a bit lower temperatures
and, in fact, as we can better see in the right panel of
Fig.~\ref{SM-V_T}, at temperatures of $T\sim$ 135, 165 and 190 GeV,
respectively, we address $\langle\phi_c(T)\rangle/T \sim 1$.

We can now compare the maximal reheating temperature as provided by
Fig.~\ref{SM-V_T} with the level contours of fixed reheating
temperature provided in the various panels of Fig.~\ref{bounce}. For a
fixed value of the Higgs mass we will first read from the right panel
of Fig.~\ref{SM-V_T} the maximal reheating temperature
$T_R^{max}$. Second in the corresponding panel of Fig.~\ref{bounce} we
will look for the contour line corresponding to the same value of the
reheating temperature for a fixed value of $\mu_{TeV}$. Then the
available region in the $(\theta,\nu)$ plane will be that to the left
and below it. For instance if we consider $m_H=140$ GeV we can see
from Fig.~\ref{SM-V_T} that $T_R^{max}\simeq 150$ GeV. Then if we fix
$\mu_{TeV}=1$ TeV and for the $N=3$ case we see in Fig.~\ref{bounce}
that there is a contour line corresponding to $T_R=0.15\, \mu_{TeV}=150$
GeV and we can easily identify the available region. Now if we
increase the Higgs mass we will obtain from Fig.~\ref{SM-V_T} a larger
value of $T_R^{max}$ and therefore going back to Fig.~\ref{bounce} the
region below and on the left of the corresponding level curve will be
bigger than the previous one. For instance if we consider now the
Higgs mass $m_H=180$ GeV we can easily see that $T_R^{max}\simeq 170$
GeV and the corresponding available region will be that below and to
the left of the contour line $T_R=0.17\, \mu_{TeV}$, and this region is
bigger than the previous one. So for a fixed value of $\mu_{TeV}$ the
heavier the Higgs boson the bigger the available region in the
parameter space.

\begin{figure}[htb]
\vspace{.7cm}
\begin{center}
\epsfig{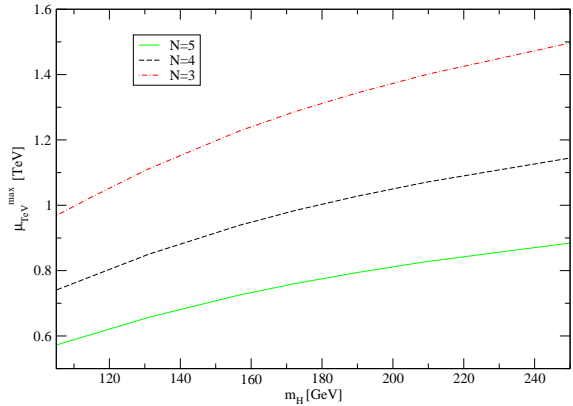}
\end{center}
\caption{Upper bound on $\mu_{TeV}$ for different values of $N$ as a function
  of the Higgs mass from the condition that $\langle\phi_c(T_R)\rangle/T_R>
  1$.}
\label{maxMUtev}
\end{figure}

Another point of view of the same condition is if we instead fix the
minimal nucleation temperature, i.e.~the maximum number of permitted
inflation, in which case the maximal reheating temperature $T_R^{max}$
given by Fig.~\ref{SM-V_T} translates into an upper bound on the value
of $\mu_{TeV}$ as can be seen in Fig.~\ref{maxMUtev} where the case of
$T_n\sim 500$ MeV, equivalent to a number of e-folds $N_e\sim 6$, has
been considered.

We can conclude that with an additional source of
CP-violation~\footnote{This extra source requires by itself extension
of the Standard Model. Some mechanisms have been proposed, {\it e.g.}
that in Ref.~\cite{Berkooz:2004kx}, where the strength of CP violation
induced by the CKM phase varies with the temperature. From the low
energy point of view another interesting possibility would be the
appearance of new CP-violating operators. For instance, as discussed
in Ref.~\cite{after29}, dimension-six operators $\sim \frac{g^2}{32
\pi^2 \mu_{TeV}^2}|H|^2 F\tilde F$ might generate the observed
baryon-to-entropy ratio. Finally an obvious possibility is to enlarge
the SM particle content in such a way as to have new sources of
CP-violation. An example is provided by the MSSM where one can have an
independent source of CP-violation {\it e.g.}~in the phase of the
Higgsino mass in which case charginos (and neutralinos) can generate
the baryon asymmetry. In our Randall-Sundrum scenario it would be
enough to enlarge the SM with the corresponding fermions (charginos
and neutralinos) as in Split Supersymmetry. However analyzing those
possible sources of CP-violation is out of the scope of the present
paper and moreover it would deserve an independent (dedicated)
investigation.} the Standard Model with a Higgs boson localized on a
brane of a warped space can generate electroweak
baryogenesis. A condition for this to happen, as we have seen, is that
the universe spans less than 26 e-folds in the supercooling previous
to the fast phase transition. This number is close to the number of
e-folds required in low-scale inflation to solve the cosmological
horizon problem~\cite{Knox:1992iy}
\begin{equation}
N_e>30+\log\frac{T_R}{\mu_{TeV}}\ .
\label{condinfl}
\end{equation}
In our case condition (\ref{condinfl}) translates into $N_e>27.7$
which is barely consistent with condition (\ref{boundTn}).  We can
even stress that in some regions of the available parameter space it
might be possible to span the number of e-folds required by
Eq.~(\ref{condinfl}) although we have not made a dedicated numerical
analysis in this paper.

\section{Conclusion and Outlook}
In this paper we have studied the supercooled electroweak phase
transition triggered by the radion in the Randall-Sundrum model at
finite temperature in the presence of a Goldberger-Wise potential
stabilizing the radion field and deduced the conditions for
electroweak baryogenesis coming from the sphaleron interaction
rates. We have concentrated in the region of the space of parameters
where the theory remains perturbative and the Randall-Sundrum metric
is not perturbed by the Goldberger-Wise field. We have estimated from
analytical approximations that the contribution of the Higgs field to
the Euclidean action is negligible (at least for small Higgs masses)
as compared to that of the radion and have worked in the approximation
of neglecting the former. Our results can be summarized as follows:

\begin{itemize}
\item
In the region of the GW parameters consistent with perturbativity
there is a wide sub-region where the phase transition is completed
provided that the parameter $N$ is not too large. In fact we have
found that for $N>6$ it is very difficult the phase transition to be
in agreement with perturbativity constraints.
\item
In the available region there is always a certain amount of inflation
which can go from a few e-folds to larger values. We have explicitly
performed our numerical analysis for a moderate maximum value of
e-folds ($N_e\leq 6$) although in some tuned regions of the
parameter space we might have the number of e-folds required to solve
the cosmological horizon problem in low-scale inflationary models
($N_e\sim 30$).
\item
In parallel with the previous issue we obtain values of the nucleation
temperatures much below typical electroweak temperatures~\footnote{As
much as we can compare our nucleation temperatures are lower than
those obtained in Ref.~\cite{Randall:2006py}.}. This makes sphalerons
to decouple inside the bubbles, as required by the condition of not
erasing any previously generated baryon asymmetry, independently of the
detailed value of the nucleation temperature.
\item
In the analyzed region bubbles percolate and there is no dangerous
production of large bubbles that could distort CMB radiation.

\item
After the phase transition the system reheats to temperatures
$T_R/\mu_{TeV}\sim0.1-0.2$. In the presence of supercooling the
reheating temperature is essentially insensitive to the precise value of the
nucleation temperature.

\item
At the reheating temperature the condition for the sphaleron to remain
decoupled translates into a lower bound on the Higgs mass. Therefore
large values of the Higgs mass are favored by electroweak baryogenesis
unlike in most models.

\item
Sphalerons outside the bubbles should be active in order to generate
baryon asymmetry in the bubbles wall. The condition for sphalerons to
be in thermal equilibrium outside the bubbles translates into an upper
bound on the number of e-folds of inflation as $N_e> 26$.
\item
In the numerical analysis we have considered the case where only the
Higgs field is localized on the IR brane. We have checked that similar
results hold for the cases where also (part of) the SM fermions are
equally localized which would basically translate into a modification
of the effective number of light degrees of freedom in the AdS-S
phase, as it can be seen from Eq.~(\ref{ultima}).
\end{itemize}

There is a number of open questions and further studies that are worth
mentioning here:

\begin{itemize}
\item
Our numerical analysis has been based on neglecting the Higgs
effective potential as compared to the GW radion potential. We believe
that this approximation is fully justified for small values of the
Higgs mass. However as electroweak baryogenesis in these models seems
to favor large over small Higgs masses it should be worthwhile to
evaluate the effect of the Higgs potential on the Euclidean action for
such heavy Higgses.
\item
For heavy Higgses the compatibility of the radion and Higgs
interactions with electroweak precision tests is an issue and should
be considered in detail~\cite{Gunion:2003px}.
\item
As the CKM phase in the Standard Model is known \cite{Gavela:1993ts} 
to be insufficient for a successful electroweak baryogenesis it should 
be essential to find out some extra source of CP violation in the considered
Randall-Sundrum model or extensions thereof.
\item
As we already mentioned we analyzed numerically cases where the number
of e-folds of inflation is moderate, $N_e\leq 6$. However in some
regions of our space of parameters we might produce a much stronger
inflation and in particular the number of e-folds required to solve
the cosmological horizon problem in low-scale inflation. It should be
interesting to explore in detail such regions since there the radion
might be identified with the inflaton field.
\item
We have considered a simplified model where the Higgs is fully
localized in the IR brane. In other models we could have similar
effects.  For instance in gauge-Higgs unification models the Higgs
field, which arises from the extra-dimensional component of a gauge
field, is exponentially localized at the IR brane and the situation
seems quite similar to the one we have studied. In Higgsless models
the gauge boson masses are provided by the IR boundary conditions and
the only sensible way in which the electroweak symmetry could be
restored at high temperatures is in the deconfined phase we
have described, in which the IR brane disappears.  In all of these
models the EW phase transition should be similar to that studied in
the present paper although dedicated studies should be worthwhile.
\end{itemize}

\subsection*{Acknowledgments}

\noindent 
Work supported in part by the European Commission under the European
Union through the Marie Curie Research and Training Networks ``Quest
for Unification" (MRTN-CT-2004-503369) and ``UniverseNet"
(MRTN-CT-2006-035863), and in part by CICYT, Spain, under contracts
FPA 2004-02012 and FPA 2005-02211.


\end{document}